\documentclass{SciPost}

\hypersetup{
    colorlinks,
    linkcolor={red!50!black},
    citecolor={blue!50!black},
    urlcolor={blue!80!black}
}

\usepackage[bitstream-charter]{mathdesign}
\usepackage{pmboxdraw}

\DeclareSymbolFont{usualmathcal}{OMS}{cmsy}{m}{n}
\DeclareSymbolFontAlphabet{\mathcal}{usualmathcal}

\fancypagestyle{SPstyle}{
\fancyhf{}
\lhead{\colorbox{scipostblue}{\bf \color{white} ~SciPost Physics Codebases }}
\rhead{\href{https://scipost.org/SciPostPhysCodeb.74}{{\bf \color{scipostdeepblue} ~SciPost Phys.\@ Codebases 74 (2026)} }}

\fancyfoot[C]{\textbf{\thepage}}
}

\usepackage{nicematrix}
\usepackage{mathabx}
\usepackage[labelformat=simple]{subcaption}

\usepackage{xcolor}
\definecolor{rwthblue}{HTML}{00549F}
\definecolor{rwthblue50}{HTML}{8EBAE5}
\definecolor{rwthblue25}{HTML}{C7DDF2}
\definecolor{rwthblue10}{HTML}{E8F1FA}
\definecolor{rwthgreen}{HTML}{57AB27}
\definecolor{rwthred}{HTML}{CC071E}
\definecolor{bash1}{gray}{0.15}
\definecolor{bash2}{gray}{.5}

\usepackage{listings}
\usepackage[listings]{tcolorbox}
\lstdefinestyle{slha}{
	linewidth=\textwidth,
	numbers=none,
	basicstyle=\footnotesize\ttfamily,
    backgroundcolor = \color{rwthblue10},
	frame=none,
    emph={Block},
    emphstyle=\color{rwthblue},
    comment=[l]{\#},
    commentstyle=\color{gray}
}
\lstdefinestyle{bashstyle}{
	linewidth=\textwidth,
	numbers=none,
	backgroundcolor=black,
	basicstyle=\footnotesize\ttfamily\color{white},
	framexleftmargin=-500pt,
	frame=none
}
\newtcblisting{bashlisting}[1][]{
    before={\\[6pt]},
    after={\par\noindent},
    arc=3pt, outer arc=3pt,
    colback=bash1,
    colframe=bash2,
    boxrule=1.5pt,
    top=-3pt,
    bottom=-3pt,
    left=-4pt,
    listing only,
    listing style=bashstyle,
    title=#1,
    }
\usepackage{wasysym}

\renewcommand{\textsc}{\bgroup\obeyspaces\helpersc}
\def\helpersc#1{\helperscii #1\relax\relax\egroup}
\def\helperscii#1{%
\ifx\relax#1\else \ifcat#1\@sptoken{} \expandafter\expandafter\expandafter\helperscii\else
\ifnum`#1=\uccode`#1 {\normalsize #1}\else {\footnotesize \uppercase{#1}}\fi \expandafter\expandafter\expandafter\helperscii\expandafter\fi\fi}

\newcommand{\history}{\texttt{history}}
\newcommand{\sushi}{\texttt{SusHi}}
\newcommand{\vhnnlo}{\texttt{vh@nnlo}}

\newcommand{\nnlojet}{\texttt{NNLOJET}}
\newcommand{\flm}{F_\mathrm{LM}}
\newcommand{\flv}{F_\mathrm{LV}}
\newcommand{\flrv}{F_\mathrm{LRV}}
\newcommand{\flvsq}{F_\mathrm{LV^2}}
\newcommand{\flvv}{F_\mathrm{LVV}}
\newcommand{\s}{\mathcal{S}}
\newcommand{\sso}{\mathcal{S}\hspace*{-6.5pt}\mathcal{S}}
\newcommand{\co}{\mathcal{C}}
\newcommand{\cco}{\mathcal{C}\hspace*{-5.2pt}\mathcal{C}}

\begin{document}

\pagestyle{SPstyle}

\begin{center}{\Large \textbf{\color{scipostdeepblue}{
{\fontfamily{lmtt}\fontseries{b}\selectfont history}: A tool for fully-differential cross sections\linebreak at next-to-next-to-leading order
}}}\end{center}

\begin{center}\textbf{
Sven Yannick Klein\textsuperscript{1$\star$} and
Lukas Simon\textsuperscript{1,2$\dagger$}
}\end{center}

\begin{center}
{\bf 1} Institute for Theoretical Particle Physics and Cosmology, RWTH Aachen University, Sommerfeldstraße 16, 52074 Aachen, Germany
\\
{\bf 2} Laboratoire de Physique Th\'eorique et Hautes Energies (LPTHE), UMR 7589, Sorbonne Universit\'e et CNRS, 4 place Jussieu, 75252 Paris Cedex 05, France
\\[\baselineskip]
$\star$ \href{mailto:yklein@physik.rwth-aachen.de}{\small yklein@physik.rwth-aachen.de}\,,\quad
$\dagger$ \href{mailto:lsimon@lpthe.jussieu.fr}{\small lsimon@lpthe.jussieu.fr}
\end{center}


\section*{\color{scipostdeepblue}{Abstract}}
\textbf{\boldmath{%
The software {\fontfamily{lmtt}\fontseries{b}\selectfont history} is designed to calculate fully-differential cross sections for colour-singlet production processes in hadronic collision up to next-to-next-to-leading order in QCD. It is based on the fully-local nested soft-collinear subtraction scheme, whose implementation is entirely process independent. This allows the program to be readily applied to arbitrary colour-singlet production processes, provided the corresponding process-dependent matrix elements are supplied. In the current release, we include matrix elements for Higgs production via gluon fusion, $pp\to H+X$, and for associated Higgs production with a heavy electroweak vector boson through the Drell-Yan-like Higgs-Strahlung mechanism, $pp\to V^\ast +X \to VH+X$, with $V\in\{W,Z\}$.
\begin{center}
    \includegraphics[width=6cm]{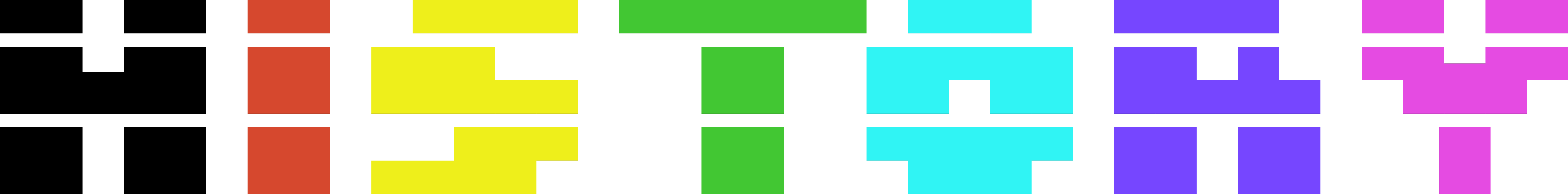}
\end{center}
}}

\vspace{\baselineskip}

\noindent\textcolor{white!90!black}{%
\fbox{\parbox{0.975\linewidth}{%
\textcolor{white!40!black}{\begin{tabular}{lr}%
  \begin{minipage}{0.6\textwidth}%
    {\small Copyright attribution to authors. \newline
    This work is a submission to SciPost Physics Codebases. \newline
    License information to appear upon publication. \newline
    Publication information to appear upon publication.}
  \end{minipage} & \begin{minipage}{0.4\textwidth}
    {\small Received Date \newline Accepted Date \newline Published Date}%
  \end{minipage}
\end{tabular}}
}}
}


\vspace{10pt}
\noindent\rule{\textwidth}{1pt}
\tableofcontents
\noindent\rule{\textwidth}{1pt}
\vspace{10pt}


\section{Introduction}\label{sec:intorduction}\noindent
The Large Hadron Collider (LHC), originally designed as a discovery machine, has evolved into a facility for precision measurements. Improved experimental methods have led to percent-level measurements and the upcoming high-luminosity phase of the LHC promises results with even greater accuracy. This progress needs to be accompanied by sufficiently accurate theoretical predictions. This is usually achieved by calculating higher orders in the strong coupling constant. As many results at NNLO are available, some processes have even been calculated at N\textsuperscript{3}LO.
Substantial progress has been made in understanding the structure of Feynman integrals, enabling the computation of increasingly complex multi-loop scattering amplitudes. Modern techniques and advanced mathematical concepts~\cite{Chetyrkin:1981qh,Binoth:2000ps,Laporta:2000dsw,Gehrmann:1999as,Tarasov:1996br,Henn_2013,vonManteuffel:2014qoa,Goncharov_2010,Muller_Stach_2013,Primo:2016ebd,Broedel_2018,Broedel_2018_2,Bourjaily:2018ycu,Smirnov:2020quc,Bourjaily:2022bwx,Gorges:2023zgv,Duhr:2024uid,Chaubey:2025adn} have led to computations of amplitudes for $2 \to 3$ scattering processes, even involving massive particles ~\cite{Abreu_2021,Chawdhry_2021,Abreu_2023,Agarwal:2021grm,Agarwal:2021vdh,Badger:2021imn,Badger:2023mgf,Abreu:2021oya,Agarwal:2023suw,DeLaurentis:2023nss,DeLaurentis:2023izi,Abreu:2021asb,Badger:2022ncb,Badger:2024sqv,Badger:2021nhg,Buonocore:2022pqq,Mazzitelli:2024ura,Buonocore:2023ljm,Becchetti:2025qlu,Badger:2021ega,Badger:2025bau,FebresCordero:2023pww,Wang:2024pmv,Agarwal:2024jyq,Devoto:2024nhl,Wang:2025kpk,Badger:2025mdi,Badger:2025ljy}. Such computations are essential for NNLO predictions. Despite these achievements and the development of many computational tools in recent years~\cite{vonManteuffel:2012np,Lee:2013mka,Duhr:2019tlz,Peraro:2019svx,Klappert:2020aqs,Hidding:2020ytt,Liu:2022chg,Armadillo:2022ugh,Magerya:2022hvj,Wu:2023upw,Heinrich:2023til,Smirnov:2023yhb,Prisco:2025wqs,Lange:2025ofh,Duhr:2026ell}, the ultimate goal of fully automated computations of NNLO scattering amplitudes currently remains out of reach.\\
Beyond amplitude calculations, a second major challenge is the automated integration over the phase space to obtain cross sections and observables for direct comparison to experimental data. Many strategies and algorithms have been developed to complete this task at NNLO~\cite{Catani_2007,Gaunt15,Ridder_2005,Heinrich_2003,Heinrich:2008si,Czakon_2010,Czakon_2011,Caola_2017,Cacciari_2015,DelDuca:2016ily,Herzog_2018,Magnea_2018,Anastasiou:2022eym}. Some of these algorithms are highly optimised for specific scenarios, such as processes with coloured particles at Born level either in the initial or in the final state but not both. Other algorithms, at least conceptually, can be used to perform the phase-space integration for arbitrarily complex NNLO processes.\\
There are two main approaches in the phase-space integration, depending on how infrared (IR) divergences are handled: non-local slicing methods and fully-local subtraction schemes. While programs for NNLO slicing methods like $q_T$ slicing~\cite{Catani_2007} and $N$-jettiness slicing~\cite{Gaunt15} have been public for many years, e.g.\@ \texttt{MATRIX}~\cite{Grazzini_2018} and \texttt{MCFM}~\cite{Boughezal_2016}, there has been a notable lack of publicly available implementations that build upon fully-local subtraction schemes. Recently, the codes \texttt{NNLOCAL}~\cite{DelDuca:2024ovc} and \nnlojet{}~\cite{huss2025nnlojet} have been released, addressing this gap, though other existing tools remain private. The former serves a proof-of-concept implementation based on the CoLoRFulNNLO subtraction scheme~\cite{Somogyi:2006cz}, albeit currently restricted to $n_f=0$. The latter employs the antenna subtraction method~\cite{Ridder_2005} and has been successfully applied to many phenomenologically relevant state-of-the-art predictions for various processes with final-state colour singlets and up to one additional jet.\\
Given that subtraction schemes such as the Catani–Seymour dipole subtraction~\cite{Catani_1997,Catani_2002} and the Frixione–Kunszt–Signer (FKS) subtraction~\cite{Frixione_1996,Frixione_1997} are known to outperform slicing methods in both precision and computational efficiency at NLO~\cite{Dittmaier_2000,Eynck_2002,BUTENSCHOEN2020115056}, an observation expected to hold also at higher orders, it is highly desirable to have more public and easily usable tools at NNLO. Hence, we release \history{} (short for \textcolor{rwthblue}{Hi}gg\textcolor{rwthblue}{s t}he\textcolor{rwthblue}{ory}) together with this article to further advance the accessibility of NNLO calculations. The software implements the fully-local nested soft-collinear (NSC) subtraction scheme~\cite{Caola_2017} for colour-singlet production processes following the description in Ref.~\cite{Caola_2019}. Here, we present a version of \history{} that can calculate cross sections for Higgs production via gluon fusion in the heavy-top limit, $pp\to H+X$, as well as for associated Higgs production with a heavy vector boson in the Drell-Yan-like production mode, $pp \to V^\ast+X\to VH+X$, where $V$ can be either a $W$ or a $Z$ boson. However, since the subtraction procedure itself is process-independent, the framework can be readily extended to any colour-singlet production process by providing the relevant matrix elements.\\
The outline of this article is as follows: In section~\ref{sec:nsc}, we review the NSC subtraction scheme and introduce the basic notation and conventions. In section~\ref{sec:ps}, we discuss the phase-space parametrisation  used in \history{}. To validate our implementation, we compare results obtained with \history{} against established codes. Specifically, for processes that are gluon-induced at LO, we present results for Higgs production via gluon fusion in section~\ref{sec:ggf}. For processes that are initiated through a quark-antiquark collision at LO, we perform a cross-check for associated Higgs boson production with a heavy vector boson in section~\ref{sec:dy}. Finally, we draw our conclusions in section~\ref{sec:conclustions}.
\section{Nested Soft-Collinear Subtraction}\label{sec:nsc}\noindent
To provide an overview of the nested soft-collinear (NSC) subtraction scheme, we briefly summarise its general structure in this section. We adhere closely to the notation and conventions of Refs.~\cite{Caola_2017,Caola_2019} to facilitate cross-referencing the more detailed descriptions in the original articles. In particular, we focus on colour-singlet production processes, for which \history{} is designed. For a generalised formulation of the NSC scheme, including processes with colourful final states at the Born level, we refer the reader to Refs.~\cite{Devoto_2023,Devoto:2025kin,Devoto:2025jql}.

\subsection{Cross Sections at Next-to-Next-to-Leading Order}\noindent
The cross section for the process $AB \to F+X$, where a final state $F$ and additional radiated partons $X$ are produced in the collision of two hadrons $A$ and $B$ can be expressed as
\begin{equation}\label{eq:sigma_had}
	\sigma\!_{AB \to F+X} = \sum_{i,j} \int_0^1 \mathrm{d}\tau\, \dfrac{\mathrm{d}\mathcal{L}_{AB}^{ij}(\tau,\mu_\mathrm{F})}{\mathrm{d}\tau} \, \hat{\sigma}_{ij \to F+X}\!\left(\tau,\mu_\mathrm{R},\mu_\mathrm{F}\right)\,.
\end{equation}
The sums over $i$ and $j$ run over all partons inside the hadrons $A$ and $B$, respectively. The parton luminosity function $\mathcal{L}_{AB}^{ij}$ is defined as a convolution integral,
\begin{equation}\label{eq:luminosity_function}
	\dfrac{\mathrm{d}\mathcal{L}_{AB}^{ij}(\tau,\mu_\mathrm{F})}{\mathrm{d}\tau} = \int_0^1 \mathrm{d}\xi_1 \mathrm{d}\xi_2 \, f_{i/A}(\xi_1,\mu_\mathrm{F}) f_{j/B}(\xi_2,\mu_\mathrm{F}) \,\delta\!\left(\xi_1 \xi_2-\tau\right) \,,
\end{equation}
where the parton distribution functions (PDFs) $f_{p/H}(\xi,\mu_\mathrm{F})$ parametrise the momentum-fraction distribution of a parton $p$ inside a hadron $H$. They encode the density of partons carrying a fraction $\xi$ of the hadron momentum, evaluated at the factorisation scale $\mu_\mathrm{F}$.\\
The partonic cross section for the process $ij \to F+X$ can be computed perturbatively as an expansion in the strong coupling constant $\alpha_s$,
\begin{equation}\label{eq:sigma_partonic}
	\hat{\sigma}_{ij \to F+X} = \hat{\sigma}_{ij \to F}^\mathrm{LO} + \hat{\sigma}_{ij \to F+X}^\mathrm{NLO} + \hat{\sigma}_{ij \to F+X}^\mathrm{NNLO} + \mathcal{O}\left(\alpha_s^{b+3}\right) \,,
\end{equation}
where the exponent $b$ is the power of the strong coupling constant of the lowest order contribution. We refer to the terms $\hat{\sigma}_{ij \to F}^\mathrm{LO}$, $\hat{\sigma}_{ij \to F+X}^\mathrm{NLO}$, and $\hat{\sigma}_{ij \to F+X}^\mathrm{NNLO}$ as leading order (LO), next-to-leading order (NLO), and next-to-next-to-leading order (NNLO) contributions, respectively. For brevity, we omitted the explicit dependence of $\hat{\sigma}_{ij \to F+X}$ on the variable $\tau$, as well as on the factorisation scale $\mu_\mathrm{F}$ and the renormalisation scale $\mu_\mathrm{R}$ in Eq.~(\ref{eq:sigma_partonic}). From now on, we fix $\mu_\mathrm{F}=\mu_\mathrm{R}=\mu$, since it is straightforward to separate the scales again in all the upcoming expressions via renormalisation-group equations.\\
At LO the final state does not contain real-radiated partons on top of the particles that we are interested in producing during the scattering process. Since this part of the cross section corresponds to its Born approximation we will call the LO cross section also the Born cross section, $\mathrm{d}\sigma_{ij \to F}^\mathrm{B}$,
\begin{equation}\label{eq:dsigma_lo}
	\mathrm{d}\hat{\sigma}_{ij \to F}^\mathrm{LO} = \mathrm{d}\hat{\sigma}_{ij \to F}^\mathrm{B}\,.
\end{equation}
At NLO the cross section receives virtual corrections, $\mathrm{d}\hat{\sigma}_{ij \to F}^\mathrm{V}$, as well as real-emission corrections, $\mathrm{d}\hat{\sigma}_{ij \to F+1}^\mathrm{R}$, with one additional massless parton in the final state compared to the Born contribution,
\begin{equation}\label{eq:dsigma_nlo}
	\mathrm{d}\hat{\sigma}_{ij \to F+X}^\mathrm{NLO} = \mathrm{d}\hat{\sigma}_{ij \to F}^\mathrm{V} + \mathrm{d}\hat{\sigma}_{ij \to F+1}^\mathrm{R} + \mathrm{d}\hat{\sigma}_{ij \to F}^\mathrm{C_{NLO}} \,.
\end{equation}
Additionally, the collinear counterterm $\mathrm{d}\hat{\sigma}_{ij \to F}^\mathrm{C_{NLO}}$ arises, which has its origin in the separation of the short- and long-distance effects during mass factorisation. This term shifts the collinear divergences of the bare PDFs into the partonic cross section, resulting in a finite definition of renormalised PDFs. While each of the three terms in Eq.~(\ref{eq:dsigma_nlo}) is divergent due to infrared (IR) singularities, their sum remains finite. This cancellation of divergences is guaranteed by the Kinoshita–Lee–Nauenberg (KLN) theorem~\cite{PhysRev.52.54,Kinoshita:1962ur,Lee:1964is} up to all orders in QCD.\\
At NNLO the cross section can be decomposed into five terms,
\begin{equation}\label{eq:dsigma_nnlo}
	\mathrm{d}\hat{\sigma}_{ij \to F+X}^\mathrm{NNLO} = \hat{\sigma}_{ij \to F}^\mathrm{VV} + \mathrm{d}\hat{\sigma}_{ij \to F+1}^\mathrm{RV} + \mathrm{d}\hat{\sigma}_{ij \to F+2}^\mathrm{RR} + \mathrm{d}\hat{\sigma}_{ij \to F}^\mathrm{C_{NNLO}} + \mathrm{d}\hat{\sigma}_{ij \to F+1}^\mathrm{C_{NNLO}} \,.
\end{equation}
Again, each term contains IR divergences, but their sum yields a finite number. In Eq.~(\ref{eq:dsigma_nnlo}), we have introduced the term with double-virtual corrections $\hat{\sigma}_{ij \to F}^\mathrm{VV}$, the real-virtual contributions $\mathrm{d}\hat{\sigma}_{ij \to F+1}^\mathrm{RV}$ with one extra parton in the final state, and the double-real emission term $\mathrm{d}\hat{\sigma}_{ij \to F+2}^\mathrm{RR}$ with two additional massless partons in comparison to the Born process. Moreover, we have two collinear counterterms $\mathrm{d}\hat{\sigma}_{ij \to F}^\mathrm{C_{NNLO}}$ and $\mathrm{d}\hat{\sigma}_{ij \to F+1}^\mathrm{C_{NNLO}}$, the former containing only the particles of the Born process, the latter one additional parton in its final state.

\subsection{Colour-Singlet Production}\noindent
To produce an arbitrary colour singlet $F_\mathrm{CS}$, we compute the partonic cross section for the process $ij \to F_\mathrm{CS}+X$. At Born level, this process is given by $i(p_1)j(p_2) \to F_\mathrm{CS}(\{k_i\})$, where we have assigned momentum $p_1$ to parton $i$, momentum $p_2$ to parton $j$, and the set of momenta $\{k_i\}=\{k_1,\dots,k_m\}$ to the $m$ particles that collectively form the colour singlet. For the LO cross section we write
\begin{equation}\label{eq:dsigma_Born}
    2\hat{s}\cdot\mathrm{d}\hat{\sigma}^\mathrm{B}_{ij \to F_\mathrm{CS}} = \flm(1_{i},2_{j}) \,.
\end{equation}
with the squared partonic centre-of-mass energy $\hat{s}=(p_1+p_2)^2$ and the compact notation
\begin{equation}\label{eq:flmqq}
    \flm(1_{i},2_{j}) = \mathcal{N}\, \mathrm{d}\Pi_\mathrm{B}^{(d)}\,\left\vert\mathcal{M}^{(0)}_{ij \to F_\mathrm{CS}}\left(p_1,p_2,\{k_i\}\right)\right\vert^2 F\!_J\left(p_1,p_2,\{k_i\}\right) \,.
\end{equation}
Eq.~(\ref{eq:flmqq}) contains the symmetry factor $\mathcal{N}$, the Lorentz-invariant Born phase-space element in $d=4-2\epsilon$ dimensions
\begin{equation}\label{eq:dLIPS_B}
    \mathrm{d}\Pi_\mathrm{B}^{(d)} = \prod_{i=1}^{m}\dfrac{\mathrm{d}^{d-1}k_i}{(2\pi)^{d-1} 2 k_i^0} \,(2\pi)^d\,\delta^{(d)}\left(p_1+p_2-\sum_{i=1}^m k_i\right) \,,
\end{equation}
the UV-renormalised\footnote{If not state otherwise, we assume all quantities to be UV renormalised in the $\overline{\text{MS}}$ scheme.} squared matrix element $\vert\mathcal{M}^{(0)}_{ij \to F_\mathrm{CS}}\vert^2$ of the tree-level process at LO, according to the first term in the expansion of the amplitude in terms of the strong coupling constant $\alpha_s$,
\begin{equation}\label{eq:amp_uvren}
    \begin{aligned}
        \mathcal{M}_{ij \to F_\mathrm{CS}}\left(p_1,p_2,\{k_i\}\right) &= \mathcal{M}_{ij \to F_\mathrm{CS}}^{(0)}\left(p_1,p_2,\{k_i\}\right)+\mathcal{M}_{ij \to F_\mathrm{CS}}^{(1)}\left(p_1,p_2,\{k_i\}\right)\\
        &\quad+\mathcal{M}_{ij \to F_\mathrm{CS}}^{(2)}\left(p_1,p_2,\{k_i\}\right)+\mathcal{O}(\alpha_s^{b/2+3}) \,,
    \end{aligned}
\end{equation}
and the measurement function $F\!_J$ which defines an IR-safe observable.\\
At NLO, we calculate the cross section for the one-loop virtual corrections to the process $i(p_1)j(p_2) \to F_\mathrm{CS}(\{k_i\})$ according to
\begin{equation}\label{eq:dsigmaV_qq}
    2\hat{s}\cdot\mathrm{d}\hat{\sigma}^\mathrm{V}_{ij \to F_\mathrm{CS}} = \flv(1_{i},2_{j}) \,,
\end{equation}
with the abbreviation
\begin{equation}\label{eq:flvqq}
\begin{aligned}
    \flv(1_{i},2_{j}) = \mathcal{N}\,\mathrm{d}\Pi_\mathrm{B}^{(d)}\,2\,\mathfrak{Re\!}\left\lbrace\mathcal{M}^{(1)}_{ij \to F_\mathrm{CS}}\left(p_1,p_2,\{k_i\}\right)\mathcal{M}^{\ast(0)}_{ij \to F_\mathrm{CS}}\left(p_1,p_2,\{k_i\}\right) \right\rbrace
    F\!_J\left(p_1,p_2,\{k_i\}\right) \,.
\end{aligned}
\end{equation}
Since the structure of the IR singularities of the function $\flv$ is known, it is convenient to rewrite it as
\begin{equation}
    \flv(1_{i},2_{j})=\dfrac{\alpha_s(\mu)}{2\pi}I_{ij}^{(1)}(\epsilon)\flm(1_{i},2_{j})+\flv^\mathrm{fin}(1_{i},2_{j})\,,
\end{equation}
with $\flv^\mathrm{fin}$ being IR finite and all singularities encoded in
\begin{equation}\label{eq:reICatani1}
    I_{ij}^{(1)}(\epsilon) = 2\,\mathfrak{Re\!}\left\lbrace{\mathcal{I}_{ij}^{(1)}(\epsilon)}\right\rbrace\,,
\end{equation}
where the Catani-Seymour insertion operator is given by~\cite{Catani_1997}
\begin{equation}\label{eq:ICatani1}
    \mathcal{I}_{ij}^{(1)}(\epsilon) = -\dfrac{e^{\epsilon\gamma_\mathrm{E}}}{\Gamma(1-\epsilon)}e^{i\pi\epsilon}\left(\dfrac{\mu^2}{\hat{s}}\right)^\epsilon\left(\dfrac{C_i}{\epsilon^2}+\dfrac{\gamma_i}{\epsilon}\right)+\mathcal{O}(\alpha_s)\,.
\end{equation}
The Casimir operator in Eq.~(\ref{eq:ICatani1}) is
\begin{equation}\label{eq:casimir_i}
    C_i=
    \begin{cases}
    C_A\,,\ \ &\text{if}\ \ i\in\{g\}\,,\\
    C_F\,,\ \ &\text{if}\ \ i\in\{q,\widebar{q}\}\,,
    \end{cases}
\end{equation}
while the anomalous dimension is given by
\begin{equation}\label{eq:gamma_i}
    \gamma_i=
    \begin{cases}
    \beta_0\,,\ \ &\text{if}\ \ i\in\{g\}\,,\\
    \dfrac{3}{2}C_F\,,\ \ &\text{if}\ \ i\in\{q,\widebar{q}\}\,,
    \end{cases}
\end{equation}
with the first coefficient of the QCD beta function
\begin{equation}
    \beta_0 = \dfrac{11}{6}C_A-\dfrac{2}{3}T_Rn_f\,,
\end{equation}
and $\gamma_\mathrm{E}$ the Euler–Mascheroni constant. The parameters of the $\mathbf{SU(3)}$ colour symmetry group with $N_c=3$ colours that are relevant for our purpose are the quadratic Casimir operator of the adjoint representation $C_A=3$, the quadratic Casimir operator of the fundamental representation $C_F=4/3$, and the index of the fundamental representation $T_R=1/2$. Additionally, we use the symbol $n_f$ to denote the number of light quark flavours, which depends on the specific theory under consideration.\\
While the IR singularities of the virtual corrections manifest themselves as explicit poles in the dimensional regulator $\epsilon$, poles from the real-radiation contributions only appear after phase-space integration. Since our goal is to perform the phase-space integration numerically, a subtraction scheme is required to isolate them and convert them to explicit poles without performing full phase-space integration. The scheme of our choice is the NSC subtraction scheme, a natural extension of the Frixione–Kunszt–Signer (FKS) subtraction scheme~\cite{Frixione_1996,Frixione_1997} from NLO to NNLO. As for the FKS subtraction scheme, in the NSC scheme each IR singularity is extracted by a separate operator. The soft operators $\mathcal{S}_k$ extract the leading singularities when parton $k$ becomes soft and therefore has vanishing energy, the operators $\mathcal{C}_{kl}$ extract the leading singularities when parton $k$ becomes collinear to parton $l$. The operators are not unique, but we define them in the same way as Ref.~\cite{Caola_2017},
\begin{equation}\label{eq:ir_operators_nlo}
    \s_k F=\lim_{p_k^0\to 0}F\,, \qquad \co_{kl}F=\lim_{\eta_{lk}\to 0}F\,,
\end{equation}
where $\eta_{lk}=(1-\cos\theta_{lk})/2$, with $\theta_{lk}$ being the angle between partons $l$ and $k$, and $F$ is an arbitrary function.\footnote{\label{fnt:lorentz_breaking}The energy $p_k^0$ and the angle $\theta_{lk}$ are chosen to be in the centre-of-mass frame of the colliding partons. This choice breaks the Lorentz invariance of the subtraction formalism.}\\
We write the cross section for the real emission $i(p_1)j(p_2) \to F_\mathrm{CS}(\{k_i\})+k(p_4)$ in Eq.~(\ref{eq:dsigma_nlo}) as
\begin{equation}\label{eq:sigmaR_fks}
\begin{aligned}
    2\hat{s}\cdot\mathrm{d}\hat{\sigma}^\mathrm{R}_{ij \to F_\mathrm{CS}+1} &= \left\langle \mathcal{\hat{O}}_\mathrm{NLO} \flm(1_{i},2_{j} \,\vert\, 4_{k}) \right\rangle + \left\langle \left( \co_{41}+\co_{42} \right) \flm(1_{i},2_{j} \,\vert\, 4_{k}) \right\rangle \\
    &\quad + \left\langle \left(1-\co_{41}-\co_{42} \right)\s_4 \flm(1_{i},2_{j} \,\vert\, 4_{k}) \right\rangle \,,
\end{aligned}
\end{equation}
where we have defined
\begin{equation}\label{eq:flm_qqg}
    \flm(1_{i},2_{j} \,\vert\, 4_{k}) = \mathcal{N}\,\mathrm{d}\Pi_\mathrm{B}^{(d)}\,\left\vert\mathcal{M}^{(0)}_{ij \to F_\mathrm{CS}+k}\left(p_1,p_2,\{k_i\},p_4\right)\right\vert^2 F\!_J\left(p_1,p_2,\{k_i\},p_4\right) \,,
\end{equation}
and the NLO subtraction operator reads
\begin{equation}\label{eq:ONLO}
	\mathcal{\hat{O}}_\mathrm{NLO} = \left( 1-\co_{41}-\co_{42} \right)\left( 1-\s_4 \right) \,.
\end{equation}
The Born phase-space element is given by\footnote{\label{fnt:phasespace}Note that the Born phase space includes the momentum-conserving $\delta$-distribution for the entire process. Thus, it differs from Eq.~(\ref{eq:dLIPS_B}). However, we assume that this does not lead to confusion and we use the same symbol.}
\begin{equation}\label{eq:dLIPS_B_NLO}
    \mathrm{d}\Pi_\mathrm{B}^{(d)} = \prod_{i=1}^{m}\dfrac{\mathrm{d}^{d-1}k_i}{(2\pi)^{d-1} 2 k_i^0} \,(2\pi)^d\,\delta^{(d)}\left(p_1+p_2-\sum_{i=1}^m k_i-p_4\right)\text{.}
\end{equation}
Additionally, we have introduced the notation for the real emission phase-space integral
\begin{equation}
    \left\langle\mathcal{\hat{O}}\flm(1_{i},2_{j} \,\vert\, 4_{k},\dots,n_{l})\right\rangle = \prod_{i=4}^{n}\int\!\!\mathcal{\hat{O}}\dfrac{\mathrm{d}^{d-1}p_i}{(2\pi)^{d-1} 2 p_i^0}\flm(1_{i},2_{j} \,\vert\, 4_{k},\dots,n_{l})
\end{equation}
with any operator $\mathcal{\hat{O}}$.\\
The first term of Eq.~(\ref{eq:sigmaR_fks}) contains the real-emission contribution and the \textit{local subtraction terms} rendering this term IR finite. The phase-space integration can be carried out numerically in $d=4$ space-time dimensions. All singularities are encoded in the second and third term, which we call \textit{integrated counterterms} since there the phase-space integration over the unresolved parton is carried out analytically. In the integrated counterterms the divergences show up as explicit poles in $\epsilon$.\\
The NLO cross section of Eq.~(\ref{eq:dsigma_nlo}) is completed by the term from the collinear renormalisation of the PDFs~\cite{Collins:1989gx}
\begin{equation}\label{eq:dsigmaC_qqg}
\begin{aligned}
	2\hat{s}\cdot\mathrm{d}\hat{\sigma}_{ij \to F_\mathrm{CS}}^\mathrm{C_{NLO}} &= \dfrac{\alpha_s(\mu)}{2\pi}\dfrac{1}{\epsilon}\sum_k\int_0^1 \!\mathrm{d}z\left\lbrack\hat{P}^{(0)}_{ki}(z) \left\langle \dfrac{\flm(z \cdot 1_{k},2_{j})}{z} \right\rangle+\hat{P}^{(0)}_{kj}(z)\left\langle \dfrac{\flm(1_{i},z \cdot 2_{k})}{z} \right\rangle\right\rbrack\,.
\end{aligned}
\end{equation}
The arguments $z \cdot 1_{k}$ or $z \cdot 2_{k}$ of the function $\flm$ indicate that the momentum $p_1$ or $p_2$ is multiplied by the factor $z$. The Altarelli–Parisi splitting functions $\hat{P}$~\cite{ALTARELLI1977298} can be found in Appendix~\ref{sec:splitting}. When plugging Eqs.~(\ref{eq:dsigmaV_qq}), (\ref{eq:sigmaR_fks}), and (\ref{eq:dsigmaC_qqg}) into Eq.~(\ref{eq:dsigma_nlo}), a pole cancellation among all the terms can be observed and the NLO cross section remains finite in the limit $\epsilon \to 0$. The explicit expression of this finite remainder is presented in Eqs.~(3.17), (3.18), and (4.27) to (4.29) of Ref.~\cite{Caola_2019}. The phase space of these process-independent formulae can be integrated fully numerically, which allows for a flexible cross-section calculation with fiducial cuts and binning with respect to arbitrary IR-safe kinematic observables. Hence, they build the core of the NLO computations with \history{}.\\
The FKS approach is well-established and used in many different tools. However, the extension to NNLO is not trivial as corrections with double-real emission of massless partons introduce new complications as we will discuss. Before we consider these contributions, we start with the first term of Eq.~(\ref{eq:dsigma_nnlo}). This term contains the double-virtual corrections of the cross section
\begin{equation}\label{eq:dsigmaVV_qq}
\begin{aligned}
    2\hat{s}\cdot\mathrm{d}\hat{\sigma}^\mathrm{VV}_{ij \to F_\mathrm{CS}} &= \bigl\langle \flvv(1_{i},2_{j}) \bigr\rangle + \bigl\langle \flvsq(1_{j},2_{j}) \bigr\rangle\,,
\end{aligned}
\end{equation}
where we distinguish the contributions $\flvv$ and $\flvsq$ which contain the interference of the two-loop amplitudes with the Born-level amplitudes and the squared one-loop amplitudes, respectively. In terms of the UV-renormalised amplitudes according to the expansion in Eq.~(\ref{eq:amp_uvren}) the functions read
\begin{equation}\label{eq:flvv_qq}
    \flvv(1_{i},2_{j}) = \mathcal{N}\, \mathrm{d}\Pi_\mathrm{B}^{(d)}\,2\,\mathfrak{Re\!}\left\lbrace\mathcal{M}^{(2)}_{ij \to F_\mathrm{CS}}(p_1,p_2,\{k_i\})\mathcal{M}^{\ast(0)}_{ij \to F_\mathrm{CS}}(p_1,p_2,\{k_i\})\right\rbrace F\!_J(p_1,p_2,\{k_i\})
\end{equation}
and
\begin{equation}\label{eq:flv2_qq}
    \flvsq(1_{i},2_{j}) = \mathcal{N}\, \mathrm{d}\Pi_\mathrm{B}^{(d)}\left\vert\mathcal{M}^{(1)}_{ij \to F_\mathrm{CS}}(p_1,p_2,\{k_i\})\right\vert^2 F\!_J(p_1,p_2,\{k_i\})\,.
\end{equation}
Like in the NLO case, the universal IR pole structure for the double-virtual correction to matrix elements is known. The finite remainder can be  defined as
\begin{equation}\label{eq:flvv_fin}
\begin{aligned}
	\flvv(1_i,2_j) &= \left(\dfrac{\alpha_s(\mu)}{2\pi}\right)^2I_{ij}^{(2)}(\epsilon) \flm(1_i,2_j) + \dfrac{\alpha_s(\mu)}{2\pi}I_{ij}^{(1)}(\epsilon) \flv(1_i,2_j) + \flvv^\mathrm{fin}(1_i,2_j)\,,
\end{aligned}
\end{equation}
where $\flvv^\mathrm{fin}$ is free of IR singularities. The colour-charge insertion operator $I_{ij}^{(1)}$ has been defined in Eq.~(\ref{eq:reICatani1}) and 
\begin{equation}\label{eq:reICatani2}
    I_{ij}^{(2)}(\epsilon) = 2\,\mathfrak{Re\!}\left\lbrace\mathcal{I}_{ij}^{(2)}(\epsilon)\right\rbrace\,,
\end{equation}
where~\cite{Catani_1998}
\begin{equation}\label{eq:ICatani2}
	\begin{aligned}
		\mathcal{I}_{ij}^{(2)}(\epsilon) &= -\dfrac{1}{2} \mathcal{I}_{ij}^{(1)}(\epsilon)\left(\mathcal{I}_{ij}^{(1)}(\epsilon)+2\dfrac{\beta_0}{\epsilon}\right) + \dfrac{\Gamma(1-2\epsilon)}{\Gamma(1-\epsilon)e^{\epsilon\gamma_\mathrm{E}}}\left(\dfrac{\beta_0}{\epsilon}+K\right) \mathcal{I}_{ij}^{(1)}(2\epsilon)\\
        &\quad+ \dfrac{1}{2\epsilon} \dfrac{e^{\epsilon\gamma_\mathrm{E}}}{\Gamma(1-\epsilon)} \mathcal{H}_{i}^{(2)}(\epsilon)\,,
	\end{aligned}
\end{equation}
with
\begin{equation}
	K = \left(\dfrac{67}{18}-\dfrac{\pi^2}{6}\right)C_A-\dfrac{10}{9}T_Rn_f
\end{equation}
and\footnote{Our definition of the $\mathcal{H}_{i}^{(2)}$ term differs from that in Eq.~(A.3) of Ref.~\cite{Caola_2019}. Consequently, the operator $\mathcal{I}_{ij}^{(2)}$ of Ref.~\cite{Caola_2019} is not scale invariant. This results in a different scale dependence for our two-loop finite remainder,\\[-18pt]
\begin{center}
    $\flvv^\mathrm{fin}(\history{})=\flvv^\mathrm{fin}(\text{Ref.~\cite{Caola_2019}})-2\,\mathcal{H}_{i}^{(2)}\!(0)\log\!\left(\dfrac{\mu^2}{\hat{s}}\right)\left(\dfrac{\alpha_s(\mu)}{2\pi}\right)^2\flm\,,$
\end{center}\vspace*{-6pt}
which is corrected through a redefinition of Eqs.~(3.62) and (4.36) in Ref.~\cite{Caola_2019} such that\\[-18pt]
\begin{center}
    $\mathrm{d}\hat{\sigma}_{(1,2),ij}^\mathrm{NNLO}(\history{})=\mathrm{d}\hat{\sigma}_{(1,2),ij}^\mathrm{NNLO}(\text{Ref.~\cite{Caola_2019}})+2\,\mathcal{H}_{i}^{(2)}\!(0)\log\!\left(\dfrac{\mu^2}{\hat{s}}\right)\left(\dfrac{\alpha_s(\mu)}{2\pi}\right)^2\dfrac{\left\langle\flm\right\rangle}{2\hat{s}}\,.$
\end{center}
}~\cite{Catani_1998,Becher_2009,Becher_2014}
\begin{equation}
\begin{aligned}
        \mathcal{H}_{g}^{(2)}(\epsilon) &= e^{-2i\pi\epsilon}\left(\dfrac{\mu^2}{\hat{s}}\right)^{2\epsilon} \left\lbrack C_A^2 \left(\dfrac{5}{12}+\dfrac{11}{144}\pi^2+\dfrac{\zeta_3}{2} \right) + C_A T_R n_f\left(-\dfrac{58}{27}-\dfrac{\pi^2}{36}\right) \right. \\
	&\quad \left. +C_F T_R n_f + \dfrac{20}{27}\left(T_R n_f\right)^2\right\rbrack \,,\\
	\mathcal{H}_{i}^{(2)}(\epsilon) &= e^{-2i\pi\epsilon}\left(\dfrac{\mu^2}{\hat{s}}\right)^{2\epsilon} \left\lbrack C_F^2 \left(-\dfrac{3}{8}+\dfrac{\pi^2}{2}-6\zeta_3 \right) + C_A C_F\left(\dfrac{245}{216}-\dfrac{23}{48}\pi^2+\dfrac{13}{2}\zeta_3\right) \right. \\
	&\quad \left. +C_F T_R n_f\left(-\dfrac{25}{54}+\dfrac{\pi^2}{12}\right)\right\rbrack \,,\quad\text{for}\ i\in\{q,\widebar{q}\}\,.
\end{aligned}
\end{equation}
The finite remainder of the squared one-loop function can be obtained via
\begin{equation}\label{eq:flvsq_fin}
\begin{aligned}
	\flvsq(1_i,2_j) &= \left(\dfrac{\alpha_s(\mu)}{2\pi}\right)^2\left\vert\mathcal{I}_{ij}^{(1)}(\epsilon)\right\vert^2 \flm(1_i,2_j) + \dfrac{\alpha_s(\mu)}{2\pi}I_{ij}^{(1)}(\epsilon) \flv^\mathrm{fin}(1_i,2_j) + \flvsq^\mathrm{fin}(1_i,2_j)\,.
\end{aligned}
\end{equation}
The next ingredient for the NNLO cross section are the real-virtual corrections. To extract the IR singularities related to the real-emitted parton, we follow the FKS procedure as done for the NLO real-emission contribution. Similar to Eq.~(\ref{eq:sigmaR_fks}), we can write
\begin{equation}\label{eq:dsigmaRV}
\begin{aligned}
    2\hat{s}\cdot\mathrm{d}\hat{\sigma}^\mathrm{RV}_{ij \to F_\mathrm{CS}+1} &= \left\langle \mathcal{\hat{O}}_\mathrm{NLO} \flrv(1_{i},2_{j} \,\vert\, 4_{k}) \right\rangle + \left\langle \left( \co_{41}+\co_{42} \right) \flrv(1_{i},2_{j} \,\vert\, 4_{k}) \right\rangle \\
    &\quad + \bigl\langle \left(1-\co_{41}-\co_{42} \right)\s_4 \flrv(1_{i},2_{j} \,\vert\, 4_{k}) \bigr\rangle\,,
\end{aligned}
\end{equation}
with
\begin{equation}\label{eq:flrv_qqg}
\begin{aligned}
    \flrv(1_{i},2_{j} \,\vert\, 4_{k}) &= \mathcal{N}\, \mathrm{d}\Pi_\mathrm{B}^{(d)}\,2\,\mathfrak{Re\!}\left\lbrace\mathcal{M}^{(1)}_{ij \to F_\mathrm{CS}+k}(p_1,p_2,\{k_i\},p_4)\mathcal{M}^{\ast(0)}_{ij \to F_\mathrm{CS}+k}(p_1,p_2,\{k_i\},p_4)\right\rbrace \\
    &\quad\times F\!_J(p_1,p_2,\{k_i\},p_4)\,.
\end{aligned}
\end{equation}
The explicit IR singularities that have their origin in the one-loop integral are universal and have the form
\begin{equation}\label{eq:FLRV_fin}
    \flrv(1_{i},2_{j} \,\vert\, 4_{k}) = \dfrac{\alpha_s(\mu)}{2\pi}\,I_{ijk}^{(1)}(\epsilon)\flm(1_{i},2_{j} \,\vert\, 4_{k})+\flrv^\mathrm{fin}(1_{i},2_{j} \,\vert\, 4_{k})
\end{equation}
where we introduce another colour-charge insertion operator
\begin{equation}
    I_{ijk}^{(1)}(\epsilon) = 2\,\mathfrak{Re\!}\left\lbrace\mathcal{I}_{ijk}^{(1)}\right\rbrace\,,
\end{equation}
with~\cite{Catani_1998}
\begin{equation}
\begin{aligned}
    \mathcal{I}_{ijk}^{(1)}(\epsilon) &= \dfrac{1}{2}e^{i\pi\epsilon}\left(\dfrac{\mu^2}{\hat{s}}\right)^{\epsilon}\left(\dfrac{C_A-2\,C_i}{\epsilon^2}+\dfrac{\chi_i-2\,\gamma_i}{\epsilon}\right) \\
    &\quad -\dfrac{1}{2}\left\lbrack\left(-\dfrac{\mu^2}{\hat{t}}\right)^{\epsilon}+\left(-\dfrac{\mu^2}{\hat{u}}\right)^{\epsilon}\right\rbrack\left(\dfrac{C_A}{\epsilon^2}+\dfrac{\chi_i +\gamma_g}{2\epsilon}\right)\,,\quad&\text{if}\ \ k\in\{g\}\,,\\
    \mathcal{I}_{ijk}^{(1)}(\epsilon) &= \dfrac{1}{2}\left(-\dfrac{\mu^2}{\hat{t}}\right)^{\epsilon}\left(\dfrac{C_A-2\,C_j}{\epsilon^2}+\dfrac{\chi_j-2\,\gamma_j}{\epsilon}\right) \\
    &\quad -\dfrac{1}{2}\left\lbrack e^{i\pi\epsilon}\left(\dfrac{\mu^2}{\hat{s}}\right)^{\epsilon}+\left(-\dfrac{\mu^2}{\hat{u}}\right)^{\epsilon}\right\rbrack\left(\dfrac{C_A}{\epsilon^2}+\dfrac{\chi_j +\gamma_g}{2\epsilon}\right)\,,\quad&\text{if}\ \ i\in\{g\}\,,\\
    \mathcal{I}_{ijk}^{(1)}(\epsilon) &= \dfrac{1}{2}\left(-\dfrac{\mu^2}{\hat{u}}\right)^{\epsilon}\left(\dfrac{C_A-2\,C_i}{\epsilon^2}+\dfrac{\chi_i-2\,\gamma_i}{\epsilon}\right) \\
    &\quad -\dfrac{1}{2}\left\lbrack e^{i\pi\epsilon}\left(\dfrac{\mu^2}{\hat{s}}\right)^{\epsilon}+\left(-\dfrac{\mu^2}{\hat{t}}\right)^{\epsilon}\right\rbrack\left(\dfrac{C_A}{\epsilon^2}+\dfrac{\chi_i +\gamma_g}{2\epsilon}\right)\,,\quad&\text{if}\ \ j\in\{g\}\,.
\end{aligned}
\end{equation}
The Casimir operators and the anomalous dimensions can be found in Eqs.~(\ref{eq:casimir_i}) and (\ref{eq:gamma_i}), respectively. Additionally we introduced
\begin{equation}\label{eq:chi_i}
    \chi_i=
    \begin{cases}
    \beta_0\,,\ \ &\text{if}\ \ i\in\{g\}\,,\\
    \dfrac{3}{2}C_A\,,\ \ &\text{if}\ \ i\in\{q,\widebar{q}\}\,.
    \end{cases}
\end{equation}
The Mandelstam variables are given according to the usual conventions,
\begin{equation}
    \hat{s}=(p_1+p_2)^2\,,\qquad\hat{t}=(p_2-p_4)^2\,,\qquad\hat{u}=(p_1-p_4)^2\,.
\end{equation}
After inserting Eq.~(\ref{eq:FLRV_fin}) into Eq.~(\ref{eq:dsigmaRV}) and integrating analytically over the $d$-dimensional phase space of the unresolved parton to obtain the integrated counterterms, all IR singularities of the real-virtual corrections are known as poles in $\epsilon$~\cite{Caola_2017}.\\
The most delicate part is the extraction of the IR singularities of the double-real emission contribution $i(p_1)j(p_2)\to F_\mathrm{CS}(\{k_i\})+k(p_4)l(p_5)$. Since new sources of IR divergences are present at NNLO, most importantly double-soft and triple-collinear singularities, we introduce the operators
\begin{equation}
    \sso F=\lim_{\substack{p_k^0\to 0\\p_l^0\to 0\\p_k^0/p_l^0=\mathrm{const.}}}\hspace{-13pt}F\,, \qquad \cco_{m} F=\lim_{\substack{\eta_{km}\to 0\\\eta_{lm}\to 0\\\eta_{km}/\eta_{lm}=\mathrm{const.}\\\eta_{kl}/\eta_{km}\not{\to} 0\, \lor\, \infty\\\eta_{kl}/\eta_{lm}\not{\to} 0\, \lor\, \infty}}\hspace{-19pt}F
\end{equation}
on top of the operators in Eq.~(\ref{eq:ir_operators_nlo}). The operator $\sso$ extracts the leading singularity when both final-state partons $k$ and $l$ become soft while their energy ratio remains constant. The operator $\cco_m$ extracts the leading singularity when both final-state partons $k$ and $l$ simultaneously become collinear to the initial-state parton $m$. In the NSC subtraction scheme, the operators can applied in a nested way and, due to colour coherence, the soft and collinear limits are commutative.\\
However, before we use these operators, we separate overlapping singularities. This is done in two steps. First, we partition the NNLO phase space into four segments by a decomposition of one according to
\begin{equation}\label{eq:partition_function}
	1 = w^{41,51} +  w^{41,52}  + w^{42,52} + w^{42,51}\,.
\end{equation}
The term proportional to the partition function $w^{41,51}$ contains only collinear singularities where parton $i$, $k$, and $l$ can become collinear to each other. The same holds for $w^{42,52}$, simply replacing initial-state parton $1$ with the second initial-state parton $2$. Since we can have triple-collinear singularities in these partitions, we call $w^{41,51}$ and $w^{42,52}$ \textit{triple-collinear partitions}. Partition $w^{41,52}$ instead allows for unregularised collinear limits only when parton $i$ and $k$  or parton $j$ and $l$ become collinear. In partition $w^{42,51}$ the roles of parton $1$ and $2$ are exchanged. Since the most singular behaviour in these partition is the simultaneous appearance of two single-collinear limits, we call them \textit{double-collinear partitions}.\\
While in the double-collinear partitions all singularities are separated, the triple-collinear partitions still contain overlapping collinear singularities, e.g.\@ in the triple-collinear limit of partition $w^{41,51}$ it is not clear in which order the partons $i$, $k$, and $l$ become collinear to each other. To resolve this issue, four \textit{sectors} are introduced in each of the two triple-collinear partitions. In partition $w^{4m,5m}$, we define the four sectors $(a)$, $(b)$, $(c)$, and $(d)$ as
\begin{equation}\label{eq:sectors}
\begin{aligned}
	1 &= \Theta\!\left( \eta_{m5} \le \frac{\eta_{m4}}{2} \right) + \Theta\!\left( \frac{\eta_{m4}}{2}  < \eta_{m5} \le \eta_{m4}\right) + \Theta\!\left( \eta_{m4} \le \frac{\eta_{m5}}{2} \right) + \Theta\!\left( \frac{\eta_{m5}}{2}  < \eta_{m4} <\eta_{m5}\right) \\
	&\equiv \Theta^{(a)}_m+\Theta^{(b)}_m+\Theta^{(c)}_m+\Theta^{(d)}_m\,.
\end{aligned}
\end{equation}
This partition of one introduces an angular ordering between the partons with a well-defined subset of non-overlapping collinear singularities in each sector.\\
In the end, we can express the double-real emission cross section as
\begin{equation}\label{eq:RR_fullyregulated}
\begin{aligned}
	2\hat{s}\cdot\mathrm{d}\hat{\sigma}^\mathrm{RR}_{ij \to F_\mathrm{CS}+2} &= \sum_{(m,n)\in\mathrm{dc}} \Bigl\langle \hat{\mathcal{O}}_\mathrm{NNLO}^{(m,n)} \flm(1_{i},2_{j} \,\vert\, 4_{k}, 5_{l}) \Bigr\rangle \\
	&\quad +\sum_{m\in\mathrm{tc}} \Bigl\langle \left(\hat{\mathcal{O}}_\mathrm{NNLO}^{(m,a)} + \hat{\mathcal{O}}_\mathrm{NNLO}^{(m,b)} + \hat{\mathcal{O}}_\mathrm{NNLO}^{(m,c)} + \hat{\mathcal{O}}_\mathrm{NNLO}^{(m,d)} \right) \flm(1_{i},2_{j} \,\vert\, 4_{k}, 5_{l}) \Bigr\rangle \\
	&\quad + \sum_{(m,n)\in\mathrm{dc}} \Bigl\langle \left( \co_{4m} + \co_{5n} - \co_{4m}\co_{5n} \right)\left( 1-\sso \right) \left( 1-\s_5 \right) w^{4m,5n} \\
	&\quad \times \flm(1_{i},2_{j} \,\vert\, 4_{k}, 5_{l}) \Bigr\rangle + \sum_{m \in\mathrm{tc}} \Bigl\langle \left( \co_{5m}\, \Theta^{(a)}_m + \co_{54}\, \Theta^{(b)}_m + \co_{4m}\, \Theta^{(c)}_m \right. \\
	&\quad \left. + \co_{45}\, \Theta^{(d)}_m \right) \left( 1-\sso \right) \left( 1-\s_5 \right) w^{4m,5m} \flm(1_{i},2_{j} \,\vert\, 4_{k}, 5_{l}) \Bigr\rangle \\
	&\quad - \sum_{m \in\mathrm{tc}} \Bigl\langle \cco_{m} \left( \co_{5m}\, \Theta^{(a)}_m + \co_{54}\, \Theta^{(b)}_m + \co_{4m}\, \Theta^{(c)}_m + \co_{45}\, \Theta^{(d)}_m \right) \left( 1-\sso \right) \\
	&\quad \times \left( 1-\s_5 \right) w^{4m,5m} \flm(1_{i},2_{j} \,\vert\, 4_{k}, 5_{l}) \Bigr\rangle + \sum_{m \in\mathrm{tc}} \Bigl\langle \cco_m \flm(1_{i},2_{j} \,\vert\, 4_{k}, 5_{l}) \Bigr\rangle \\
	&\quad + \Bigl\langle \left(1-\sso \right) \s_5 \flm(1_{i},2_{f} \,\vert\, 4_{k}, 5_{l}) \Bigr\rangle + \Bigl\langle \sso \flm(1_{i},2_{j} \,\vert\, 4_{k}, 5_{l}) \Bigr\rangle \,,
\end{aligned}
\end{equation}
with the abbreviation
\begin{equation}\label{eq:flm_ijkl}
\begin{aligned}
    \flm(1_{i},2_{j} \,\vert\, 4_{k}, 5_{l}) &= \mathcal{N}\, \mathrm{d}\Pi_\mathrm{B}^{(d)}\,\vert\mathcal{M}^{(0)}_{ij \to F_\mathrm{CS}+kl}(p_1,p_2,\{k_i\},p_4,p_5)\vert^2 \\
    &\quad\times F\!_J(p_1,p_2,\{k_i\},p_4,p_5)\,\Theta_{kl}(p^0_4,p^0_5) \,,
\end{aligned}
\end{equation}
the NNLO subtraction operators in the double-collinear partitions
\begin{equation}\label{eq:nnlooperator_dc}
	\hat{\mathcal{O}}_\mathrm{NNLO}^{(m,n)} = \left( 1-\co_{4m} \right)\left( 1-\co_{5n} \right)\left( 1-\sso \right)\left( 1-\s_5 \right) w^{4m,5n} \,,
\end{equation}
and the NNLO subtraction operators in the triple-collinear partitions
\begin{equation}\label{eq:nnlooperator_tc}
\begin{aligned}
    \hat{\mathcal{O}}_\mathrm{NNLO}^{(m,a)} &= \left( 1-\cco_{m} \right)\left( 1-\co_{5m} \right)\left( 1-\sso \right)\left( 1-\s_5 \right) w^{4m,5m} \Theta^{(a)}_m \,,\\
	\hat{\mathcal{O}}_\mathrm{NNLO}^{(m,b)} &= \left( 1-\cco_{m} \right)\left( 1-\co_{54} \right)\left( 1-\sso \right)\left( 1-\s_5 \right) w^{4m,5m} \Theta^{(b)}_m \,,\\
	\hat{\mathcal{O}}_\mathrm{NNLO}^{(m,c)} &= \left( 1-\cco_{m} \right)\left( 1-\co_{4m} \right)\left( 1-\sso \right)\left( 1-\s_5 \right) w^{4m,5m} \Theta^{(c)}_m \,,\\
	\hat{\mathcal{O}}_\mathrm{NNLO}^{(m,d)} &= \left( 1-\cco_{m} \right)\left( 1-\co_{45} \right)\left( 1-\sso \right)\left( 1-\s_5 \right) w^{4m,5m} \Theta^{(d)}_m \,.
\end{aligned}
\end{equation}
The sums in Eq.~(\ref{eq:RR_fullyregulated}) run over the sets of the double-collinear partitions $\mathrm{dc}=\{(1,2),(2,1)\}$ and the triple-collinear partitions $\mathrm{tc}=\{1,2\}$. As already discussed in the NLO case, Eq.~(\ref{eq:flm_ijkl}) contains the Born phase-space element with the momentum conserving $\delta$-distribution for the entire process, cf.\@ footnote~\ref{fnt:phasespace},
\begin{equation}\label{eq:LipsB_NNLO}
    \mathrm{d}\Pi_\mathrm{B}^{(d)} = \prod_{i=1}^{m}\dfrac{\mathrm{d}^{d-1}k_i}{(2\pi)^{d-1} 2 k^0_i} \,(2\pi)^d\,\delta^{(d)}\left(p_1+p_2-\sum_{i=1}^{m}k_i-p_4-p_5\right) \,.
\end{equation}
In addition, we have used symmetry properties to introduce the energy ordering
\begin{equation}\label{eq:energy_ordering}
    \Theta_{kl}(p^0_4,p^0_5) = \begin{cases}
        2\,\Theta(p^0_4-p^0_5)\,,\quad&\text{if}\ \ kl\in\{g\!g\}\,,\\
        1\,,\quad&\text{else}\,,
    \end{cases}
\end{equation}
for final states with two real-emitted gluons. The energy ordering allows us to simplify the regulation of single-soft singularities such that we only need the soft operator $\s_5$ in Eqs.~(\ref{eq:nnlooperator_dc}) and (\ref{eq:nnlooperator_tc}), but no operator $\s_4$.\footnote{If just real-emitted quarks and/or antiquarks are present in the final state, quantum number conservation does not allow for a single parton becoming soft. Then, we have $\s_m\flm=0$, for $m\in\{4,5\}$. If we have a quark or an antiquark and a gluon in the final state, we always choose the (anti)quark to have index $4$ and the gluon is assigned to index $5$, so that in this scenario $\s_4\flm=0$. In conclusion, there is no case where the soft limit related to $\s_4$ appears.} The local subtraction terms in Eq.~(\ref{eq:RR_fullyregulated}) can be constructed from the IR limits described in Refs.~\cite{Catani_1999,Catani_2000} and the integrated counterterms have been computed analytically in Refs.~\cite{Caola_2017,Caola_2018_2,Delto:2019asp}.\\
In a last step, the double-virtual, the real-virtual, and the double-real contribution are combined with the terms from the collinear PDF renormalisation. The missing pieces to complete Eq.~(\ref{eq:dsigma_nnlo}) read
\begin{equation}\label{eq:dsigma_cnnlo_0}
    \begin{aligned}
        2\hat{s}&\cdot\mathrm{d}\hat{\sigma}_{ij \to F_\mathrm{CS}}^\mathrm{C_{NNLO}} \\
        &= \left(\dfrac{\alpha_s(\mu)}{2\pi}\right)^2\Bigg\lbrace\sum_{k}\int_0^1 \!\mathrm{d}z\left\lbrack\sum_{l}\dfrac{\left(\hat{P}^{(0)}_{li}\otimes\hat{P}^{(0)}_{kl}\right)\!(z)}{2\epsilon^2}-\dfrac{\beta_0 \hat{P}^{(0)}_{ki}(z)}{2\epsilon^2}+\dfrac{\hat{P}^{(1)}_{ki}\!(z)}{2\epsilon}\right\rbrack\left\langle \dfrac{\flm(z \cdot 1_{k},2_{j})}{z} \right\rangle\\
        &\quad+\sum_{k}\int_0^1 \!\mathrm{d}z\left\langle \dfrac{\flm(1_{i},z \cdot 2_{k})}{z} \right\rangle\left\lbrack\sum_{l}\dfrac{\left(\hat{P}^{(0)}_{lj}\otimes\hat{P}^{(0)}_{kl}\right)\!(z)}{2\epsilon^2}-\dfrac{\beta_0 \hat{P}^{(0)}_{kj}(z)}{2\epsilon^2}+\dfrac{\hat{P}^{(1)}_{kj}\!(z)}{2\epsilon}\right\rbrack\\
        &\quad+\dfrac{1}{\epsilon^2}\sum_{k,l}\int_0^1 \!\mathrm{d}z\,\mathrm{d}\bar{z}\,\hat{P}^{(0)}_{ki}(z)\left\langle \dfrac{\flm(z \cdot 1_{k},\bar{z} \cdot 2_{l})}{z\bar{z}} \right\rangle\hat{P}^{(0)}_{lj}(\bar{z})\Bigg\rbrace\\
        &\quad+\dfrac{\alpha_s(\mu)}{2\pi}\dfrac{1}{\epsilon}\sum_{k}\int_0^1 \!\mathrm{d}z\left\lbrack\hat{P}^{(0)}_{ki}(z)\left\langle \dfrac{\flv(z \cdot 1_{k},2_{j})}{z} \right\rangle+\left\langle \dfrac{\flv(1_{i},z \cdot 2_{k})}{z} \right\rangle\hat{P}^{(0)}_{kj}(z)\right\rbrack
    \end{aligned}
\end{equation}
and
\begin{equation}
    \begin{aligned}
        2\hat{s}\cdot\mathrm{d}\hat{\sigma}_{ij \to F_\mathrm{CS}+1}^\mathrm{C_{NNLO}} &= \dfrac{\alpha_s(\mu)}{2\pi}\dfrac{1}{\epsilon}\sum_{l}\int_0^1 \!\mathrm{d}z\left\lbrack\hat{P}^{(0)}_{li}(z)\left\langle \dfrac{\flm(z \cdot 1_{l},2_{j}  \,\vert\, 4_{k})}{z} \right\rangle\right.\\
        &\quad\left.+\left\langle \dfrac{\flm(1_{i},z \cdot 2_{l}  \,\vert\, 4_{k})}{z} \right\rangle\hat{P}^{(0)}_{lj}(z)\right\rbrack\,.
    \end{aligned}
\end{equation}
All required Altarelli–Parisi splitting functions are listed in Appendix~\ref{sec:splitting}. The convolution symbol appearing in Eq.~(\ref{eq:dsigma_cnnlo_0}) is defined as
\begin{equation}\label{eq:convolution}
 (f \otimes g)(z) = \int_0^1\! \mathrm{d}x\, \mathrm{d}y f(x)g(y) \delta(z-xy) \,.
\end{equation}
Finally, it can be shown that all poles in $\epsilon$ cancel, proving that Eq.~(\ref{eq:dsigma_nnlo}) is indeed free of IR singularities. The expression for the finite remainder can be taken from sections 3 and 4 of Ref.~\cite{Caola_2019}.

\section{Phase-Space Parametrisation}\label{sec:ps}\noindent
Since the finite remainder is free of singularities after the analytic integrations of the counterterms, all remaining integrals to evaluate Eq.~(\ref{eq:sigma_had}) can be handled numerically. However, at NNLO the phase-space parametrisation in accordance with the subtraction scheme is not trivial. First introduced in Refs.~\cite{Czakon_2010,Czakon_2011}, we present in this section the parametrisation that is used in \history{}. Like in the last section, we still follow the notation used in Ref.~\cite{Caola_2019}. Here, we focus only on the $d=4$-dimensionsional representation that is required for the Monte-Carlo integration, for details on the $d=4-2\epsilon$ dimensional form we refer the reader to the original literature.\\
Due to the choice of the subtraction terms, we need to work in the collisional partonic centre-of-mass frame, cf.\@ footnote~\ref{fnt:lorentz_breaking}. The incoming parton momenta $p_1$ and $p_2$ can be expressed in terms of partonic centre-of-mass energy $\sqrt{\hat{s}}$ as
\begin{equation}\label{eq:mom_param_in}
\begin{aligned}
  p_{1} &= \dfrac{\sqrt{\hat{s}}}{2} \bigl( 1,0,0, 1 \bigr)^\mathrm{T} \,,\\
  p_{2} &= \dfrac{\sqrt{\hat{s}}}{2} \bigl( 1,0,0,- 1 \bigr)^\mathrm{T} \,.
\end{aligned}
\end{equation}
Before we describe the parametrisation of the final-state momenta, it is convenient to re-express the integration over the Bjorken momenta fractions in Eq.~(\ref{eq:luminosity_function}) in terms of the partonic centre-of-mass energy and the rapidity of the partonic centre-of-mass frame with respect to the hadronic centre-of-mass frame $y$. We use the substitution
\begin{equation}\label{eq:xi2}
 \xi_1 = \sqrt{\dfrac{\,\hat{s}\,}{s}}\,e^y \,, \qquad
 \xi_2 = \sqrt{\dfrac{\,\hat{s}\,}{s}}\,e^{-y} \,,
\end{equation}
to obtain
\begin{equation}\label{eq:bjorken_integration}
\mathrm{d}\xi_1 \mathrm{d}\xi_2 = \dfrac{1}{s} \mathrm{d}\hat{s}\,\mathrm{d}y \,,
\end{equation}
where $\sqrt{s}$ is the centre-of-mass energy of the hadronic collision.\\
This at hand, we consider the final-state phase space. The Born phase-space element of Eq.~(\ref{eq:LipsB_NNLO}) can be reshaped by introduction of a one
\begin{equation}\label{eq:LipsB_NNLO_new}
\begin{aligned}
    \mathrm{d}\Pi_\mathrm{B}^{(4)} &= \mathrm{d}^{4}p_3\,\delta^{(4)}\left(p_1+p_2-p_3-p_4-p_5\right)\prod_{i=1}^{m}\dfrac{\mathrm{d}^{3}k_i}{(2\pi)^{3} 2 k^0_i} \,(2\pi)^4\,\delta^{(4)}\left(p_3-\sum_{i=1}^{m}k_i\right)\\
    &=  \dfrac{\mathrm{d}q^2}{2\pi}\dfrac{\mathrm{d}^{3}p_3}{(2\pi)^{3} 2 p^0_3}\,(2\pi)^4\,\delta^{(4)}\left(p_1+p_2-p_3-p_4-p_5\right)\\
    &\quad\times\prod_{i=1}^{m}\dfrac{\mathrm{d}^{3}k_i}{(2\pi)^{3} 2 k^0_i} \,(2\pi)^4\,\delta^{(4)}\left(p_3-\sum_{i=1}^{m}k_i\right)\\
    &=  \mathrm{d}q^2\dfrac{\hat{s}}{q^2}\,\delta\!\left(\hat{s}-\dfrac{q^2}{\hat{p}_3^2}\right)\prod_{i=1}^{m}\dfrac{\mathrm{d}^{3}k_i}{(2\pi)^{3} 2 k^0_i} \,(2\pi)^4\,\delta^{(4)}\left(p_3-\sum_{i=1}^{m}k_i\right)\,,
    \end{aligned}
\end{equation}
where $q^2=p_3^2$ is the square of the colour-singlet invariant mass and $\hat{p}_3^2=p_3^2/\hat{s}$ is an auxiliary quantity whose analytic expression we will derive later. For now, we can assume to know its shape and eliminate the first $\delta$-distribution in the last line by the $\hat{s}$ integration from Eq.~(\ref{eq:bjorken_integration}). Then, the Born phase-space integration is reduced to an integration over the invariant mass $q^2$ with the boundaries
\begin{equation}
    q^2 \in \left\lbrack\left(\sum_{i=1}^m \sqrt{k_i^2}\right)^2,s\right)
\end{equation}
and the phase space of the colour singlet constituents whose $\delta$-distribution is now independent of the real-emitted partons. After generating the invariant mass $q^2$, the momenta of the colour-singlet constituents can be generated with any phase-space point generator in their common centre-of-mass frame and later, when also the momenta of the real-emitted partons are known (this will be described in the next paragraph), they can be boosted into the partonic centre-of-mass frame.\\
In terms of Euler angles we write the momenta of the real-emitted partons as
\begin{equation}\label{eq:mom_param_real}
\begin{aligned}
  p_{4} &= E_4 \bigl(1,\cos(\phi_4)\sin(\theta_4),\sin(\phi_4)\sin(\theta_4),\cos(\theta_4)\bigr)^\mathrm{T} \,,\\
  p_{5} &= E_5 \bigl(1,\cos(\phi_5)\sin(\theta_5),\sin(\phi_5)\sin(\theta_5),\cos(\theta_5)\bigr)^\mathrm{T} \,.
\end{aligned}
\end{equation}
The energies can be further decomposed into
\begin{equation}\label{eq:energy_param}
	E_4 = x_1 \dfrac{\sqrt{\hat{s}}}{2} \,, \qquad E_5 = x_1 x_2 \dfrac{\sqrt{\hat{s}}}{2} \,,
\end{equation}
with $x_1, x_2 \in \lbrack 0,1)$. Here, we consider the case with energy ordering, cf.\@ Eq.~(\ref{eq:energy_ordering}). The scenario without energy ordering can be obtained by the substitution $x_2 \to x_2/x_1$. Based on this, we use momentum conservation $q^2=(p_1+p_2-p_4-p_5)^2$ to deduce
\begin{equation}\label{eq:shat_param}
	\hat{s} = \dfrac{p_3^2}{\hat{p}_3^2} =  \dfrac{q^2}{1-x_1-x_1 x_2 + x_1^2 x_2 \eta_{45}} \,.
\end{equation} 
In the \textit{double-collinear partitions} it is convenient to parametrise the polar angles according to
\begin{equation}\label{eq:angle_para_dc}
    \cos(\theta_4) = 1-2 x_3 \,, \qquad \cos(\theta_5) = 1-2 x_4\,,
\end{equation}
where $x_3, x_4 \in \lbrack 0,1)$. We keep the usual definition of the azimuthal angles in the range $\phi_4,\phi_5\in[0,2\pi)$. Then, the phase-space measure for the two additionally emitted partons becomes\footnote{Note that we explicitly combine the Heaviside function of Eq.~(\ref{eq:energy_ordering}) with the real-emission phase-space integration in Eq.~(\ref{eq:dPS_real_eo}) to underline the fact that we work in the scenario with energy ordering.}
\begin{equation}\label{eq:dPS_real_eo}
	\dfrac{\mathrm{d}^{3}p_4}{(2\pi)^{3} 2 p_4^0}\dfrac{\mathrm{d}^{3}p_5}{(2\pi)^{3} 2 p_5^0} \Theta(p^0_4-p^0_5) = \dfrac{\hat{s}^2}{16(2\pi)^4} x_1^3 x_2\, \mathrm{d}x_1 \mathrm{d}x_2 \mathrm{d}x_3 \mathrm{d}x_4 \dfrac{\mathrm{d}\phi_4}{2\pi} \dfrac{\mathrm{d}\phi_5}{2\pi} \,.
\end{equation}
Moreover, the IR operator can be rewritten in the simplified form
\begin{equation}\label{eq:IRlimits_dc}
\begin{aligned}
    &\s_5 F = \lim_{x_2 \to 0} F \,,\quad & \sso F = \lim_{x_1 \to 0} F \,,\\
    &\co_{41} F = \lim_{x_3 \to 0} F \,,\quad & \co_{42} F = \lim_{x_3 \to 1} F\,, \\
    &\co_{51} F = \lim_{x_4 \to 0} F \,,\quad & \co_{52} F = \lim_{x_4 \to 1} F \,.
\end{aligned}
\end{equation}
Since Euler angles are not suited to parametrise the angles in the \textit{triple-collinear partitions}, it is crucial to apply a proper parametrisation to each phase-space region depending on the present IR singularities, i.e.\@ we adopt the parametrisation proposed in Refs.~\cite{Czakon_2010,Czakon_2011}, which is a key element of the sector-improved residue subtraction scheme and the NSC subtraction scheme. The parametrisation depends on the sector. Considering partition $w^{4m,5m}$, the angels can be expressed as
\begin{equation}\label{eq:tc_polar}
\begin{aligned}
	&\text{Sector }(a): \quad   &\eta_{m4} &= x_3 \,,&\eta_{m5} &= \dfrac{x_3 x_4}{2} \,,\\
	&\text{Sector }(b): \quad   &\eta_{m4} &= x_3 \,,&\eta_{m5} &= x_3\left(1-\dfrac{x_4}{2}\right) \,,\\
	&\text{Sector }(c): \quad   &\eta_{m4} &= \dfrac{x_3 x_4}{2} \,,&\eta_{m5} &= x_3 \,,\\
	&\text{Sector }(d): \quad   &\eta_{m4} &= x_3\left(1-\dfrac{x_4}{2}\right) \,,\quad   &\eta_{m5} &= x_3 \,,
\end{aligned}
\end{equation}
with $x_3,x_4 \in [0,1)$. For the angle between the final-state partons, we use
\begin{equation}\label{eq:eta45_param}
\begin{aligned}
	&\text{Sectors }(a)\texttt{\&}(c): \quad   &\eta_{45} &= \dfrac{x_3(1-x_4/2)^2}{N_\mathrm{F}(x_3,x_4/2,\lambda)} \,, \\
	&\text{Sectors }(b)\texttt{\&}(d): \quad   &\eta_{45} &= \dfrac{x_3(x_4/2)^2}{N_\mathrm{F}(x_3,1-x_4/2,\lambda)} \,,
\end{aligned}
\end{equation}
where $\lambda \in \lbrack 0,1)$ and
\begin{equation}
	N_\mathrm{F}(x_3,x_4,\lambda)=1+x_4(1-2x_3)-2(1-2\lambda)\sqrt{x_4(1-x_3)(1-x_3 x_4)} \,.
\end{equation}
With that, also the relative azimuthal angle $\phi_{45}$ is fixed, where $\phi_5=\phi_4+\phi_{45}$. We find
\begin{equation}\label{eq:phi45_ac}
\begin{aligned}
	&\text{Sectors }(a)\texttt{\&}(c): \quad &\sin(\phi_{45}) &= \dfrac{2\sqrt{\lambda(1-\lambda)}(1-x_4/2)}{N_\mathrm{F}(x_3,x_4/2,\lambda)} \,,\\
       &\text{Sectors }(b)\texttt{\&}(d): \quad &\sin(\phi_{45}) &= \dfrac{2\sqrt{\lambda(1-\lambda)}(x_4/2)}{N_\mathrm{F}(x_3,1-x_4/2,\lambda)}\,,
\end{aligned}
\end{equation}
and
\begin{equation}
    \cos(\phi_{45}) = \pm \sqrt{1-\sin^2(\phi_{45})}\,,
\end{equation}
where the negative sign for the cosine is selected if
\begin{equation}
\begin{aligned}
	&\text{Sectors }(a)\texttt{\&}(c): \quad &\dfrac{2(1-x_4/2)^2}{N_\mathrm{F}(x_3,x_4/2,\lambda)}&>2+x_4(1-2x_3) \,,\\
        &\text{Sectors }(b)\texttt{\&}(d): \quad &\dfrac{2(x_4/2)^2}{N_\mathrm{F}(x_3,1-x_4/2,\lambda)}&>2+(2-x_4)(1-2x_3) \,.
\end{aligned}
\end{equation}
The phase space for the sectors $(a)$ and $(c)$ is given by
\begin{equation}
\begin{aligned}\label{eq:dPS_sector_ac}
	\dfrac{\mathrm{d}^{3}p_4}{(2\pi)^{3} 2 p_4^0}\dfrac{\mathrm{d}^{3}p_5}{(2\pi)^{3} 2 p_5^0} \Theta(p^0_4-p^0_5)\, \Theta^{(s)}_{m} &= \dfrac{\hat{s}^2}{16(2\pi)^4} \dfrac{x_1^3 x_2 x_3(1-x_4/2)}{2N_\mathrm{F}(x_3,x_4/2,\lambda)}\\
	&\quad \times \mathrm{d}x_1 \mathrm{d}x_2 \mathrm{d}x_3 \mathrm{d}x_4 \dfrac{\mathrm{d}\phi_4}{2\pi} \dfrac{\mathrm{d}\lambda}{\pi\sqrt{\lambda(1-\lambda)}} \,,
\end{aligned}
\end{equation}
with $s\in\{a,c\}$ and the azimuthal angle $\phi_4\in[0,2\pi)$. The phase space for the sectors $(b)$ and $(d)$ is given by
\begin{equation}\label{eq:dPS_sector_bd}
\begin{aligned}
	\dfrac{\mathrm{d}^{3}p_4}{(2\pi)^{3} 2 p_4^0}\dfrac{\mathrm{d}^{3}p_5}{(2\pi)^{3} 2 p_5^0} \Theta(p^0_4-p^0_5)\, \Theta^{(s)}_{m} &= \dfrac{\hat{s}^2}{16(2\pi)^4} \dfrac{x_1^3 x_2 x_3(x_4/2)}{2N_\mathrm{F}(x_3,1-x_4/2,\lambda)}\\
	&\quad \times \mathrm{d}x_1 \mathrm{d}x_2 \mathrm{d}x_3 \mathrm{d}x_4 \dfrac{\mathrm{d}\phi_4}{2\pi} \dfrac{\mathrm{d}\lambda}{\pi\sqrt{\lambda(1-\lambda)}} \,,
\end{aligned}
\end{equation}
with $s\in\{b,d\}$ and $\phi_4\in[0,2\pi)$.\\
To avoid integrable singularities and to improve numerical stability it is convenient to rewrite
\begin{equation}
	\lambda = \sin^2\left(\dfrac{\pi}{2}x_{\lambda}\right) \,, \quad \text{with} \quad x_{\lambda}\in [0,1) \,,
\end{equation}
in Eqs.~(\ref{eq:dPS_sector_ac}) and (\ref{eq:dPS_sector_bd}) so that
\begin{equation}
	\dfrac{\mathrm{d}\lambda}{\pi\sqrt{\lambda(1-\lambda)}} = \mathrm{d}x_{\lambda} \,.
\end{equation}
The simplified soft limits are unchanged with respect to the double-collinear partitions. The relevant collinear limits are
\begin{equation}\label{eq:IRlimits_tc}
\begin{aligned}
    & &\s_5 F = \lim_{x_2 \to 0} F \,,\quad & \sso F = \lim_{x_1 \to 0} F \,,\\
    &\text{Sectors }(a): \quad&\co_{5m} F = \lim_{x_4 \to 0} F \,, \quad & \cco_{m} F = \lim_{x_3 \to 0} F \,,\\
    &\text{Sectors }(b): \quad&\co_{54} F = \lim_{x_4 \to 0} F \,, \quad & \cco_{m} F = \lim_{x_3 \to 0} F \,,\\
    &\text{Sectors }(c): \quad&\co_{4m} F = \lim_{x_4 \to 0} F \,, \quad & \cco_{m} F = \lim_{x_3 \to 0} F \,,\\
    &\text{Sectors }(d): \quad&\co_{45} F = \lim_{x_4 \to 0} F \,, \quad & \cco_{m} F = \lim_{x_3 \to 0} F \,.
\end{aligned}
\end{equation}
Finally, all momenta can be expressed in terms of the new integration variables. For the fully-resolved phase-space point we get
\begin{equation}
	p_i(q^2,x_1,x_2,x_3,x_4,\phi_4,\phi_5)
\end{equation}
in the double-collinear partitions and
\begin{equation}
	p_i(q^2,x_1,x_2,x_3,x_4,\phi_4,\lambda)
\end{equation}
in the triple-collinear partitions for all external partons $i\in\{1,2,3,4,5\}$. The related phase-space point in any IR limit that is described by the operator $\mathcal{\hat{O}}$ can be obtained by application of the IR operators according to Eq.~(\ref{eq:IRlimits_dc})
\begin{equation}
	p_i^\mathcal{\hat{O}}=\mathcal{\hat{O}}\,p_i(q^2,x_1,x_2,x_3,x_4,\phi_4,\phi_5)
\end{equation}
in the double-collinear partitions and according to Eq.~(\ref{eq:IRlimits_tc})
\begin{equation}
	p_i^\mathcal{\hat{O}}=\mathcal{\hat{O}}\,p_i(q^2,x_1,x_2,x_3,x_4,\phi_4,\lambda)
\end{equation}
in the triple-collinear partitions.
\section{Higgs-Boson Production in Gluon Fusion}\label{sec:ggf}
\begin{figure}
    \begin{subfigure}[b]{0.49\textwidth}\centering
        \includegraphics[height=3.5cm]{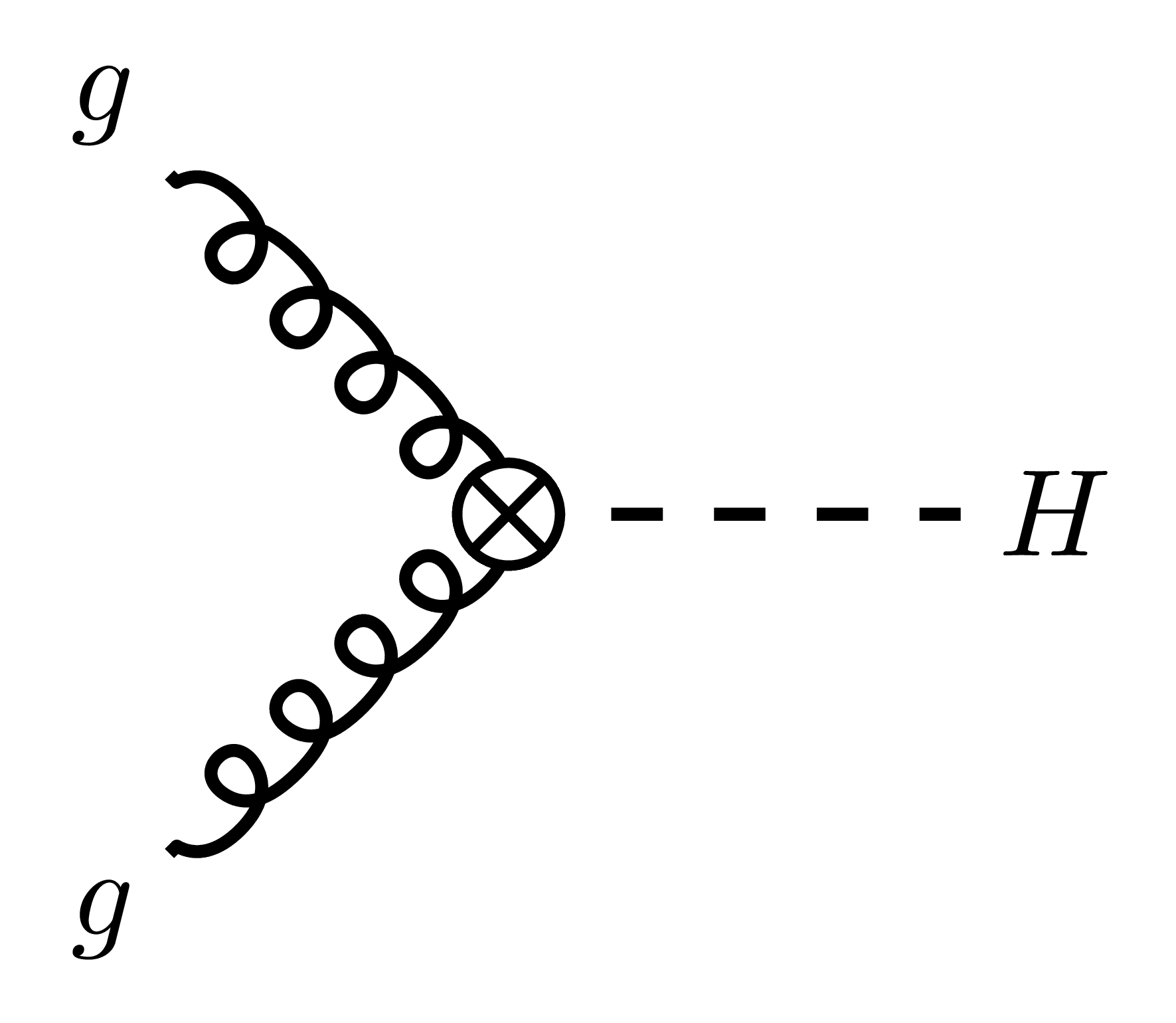}
        \caption{}\label{fig:ggf_lo}
    \end{subfigure}
    \begin{subfigure}[b]{0.49\textwidth}\centering
        \includegraphics[height=3.5cm]{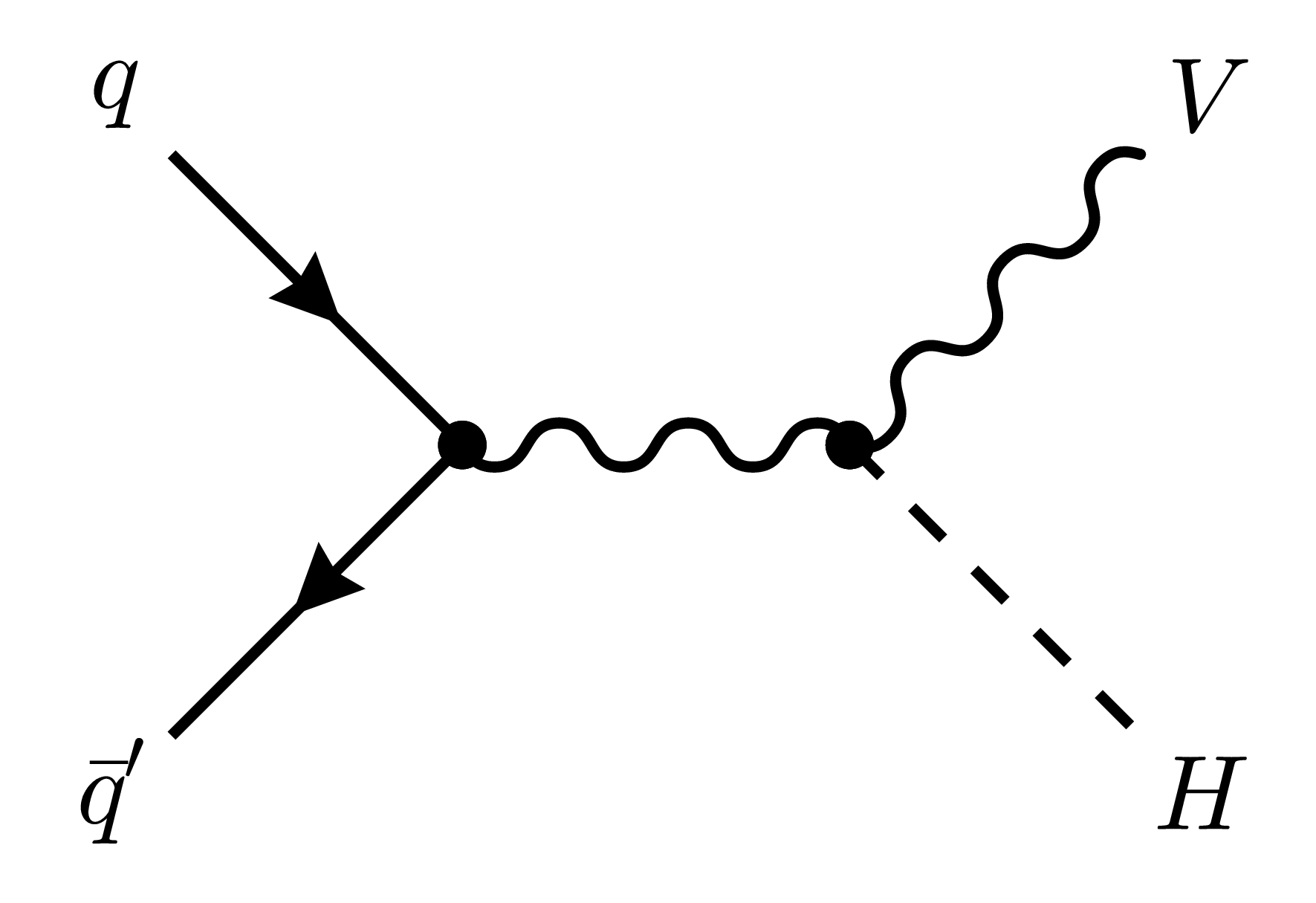}
        \caption{}\label{fig:dy_lo}
    \end{subfigure}
    \caption{Feynman diagrams for (a) Higgs production via gluon fusion in the heavy-top limit and (b) Drell-Yan-like production of a Higgs boson in association with a massive weak gauge boson $V\in\{W,Z\}$. The crossed vertex represents an effective vertex where a Higgs boson couples to a pair of gluons through a loop with infinitely heavy top quarks. These Feynman diagrams are created with \texttt{FeynGame}~\cite{feyngame,feyngame2,feyngame3}.}
    \label{fig:feynman}
\end{figure}

\begin{figure}
    \centering
    \input{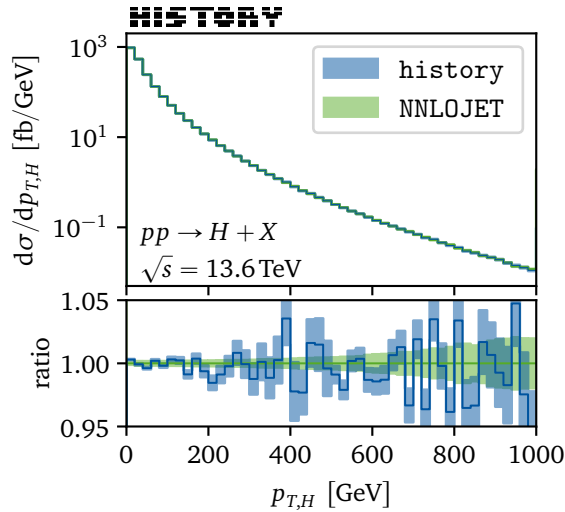}
    \caption{Transverse-momentum distribution of the Higgs boson at NNLO QCD precision. The upper panel shows the spectra computed with \history{} (blue) and \nnlojet{} (green). The lower panel illustrates their ratios normalised to the \nnlojet{} distribution. The shaded bands indicate the statistical Monte-Carlo uncertainties.}
    \label{fig:Higgs}
\end{figure}

\noindent
Since colour-singlet production in hadronic collisions can only be initiated at Born level through either gluon fusion or quark-antiquark annihilation, validating our implementation with one representative process from each class provides a non-trivial and comprehensive test of its generality. In this section, we consider the gluon-fusion case and study Higgs-boson production in proton–proton collisions, $pp\to H+X$, within the heavy-top limit. At leading order, Higgs production proceeds exclusively via the partonic channel $gg\to H$ if couplings of the light quarks in the proton to the Higgs are neglected. The corresponding Feynman diagram is shown in Fig.~\ref{fig:ggf_lo}.\\
\noindent
Higher-order QCD corrections in the heavy-top limit have been investigated extensively over the past decades. The NLO corrections were computed in the early 1990s~\cite{Dawson:1990zj,Djouadi:1991tka}, followed by the NNLO cross section in the early 2000s~\cite{Harlander:2002wh,Ravindran:2002dc,Anastasiou:2002yz}. The perturbative expansion has since been extended to N\textsuperscript{3}LO accuracy~\cite{Anastasiou:2015vya,Anastasiou:2016cez,Mistlberger:2018etf}. However, achieving the level of precision required by current experimental measurements necessitates going beyond the heavy-top approximation. Finite quark-mass effects can induce sizeable corrections and must therefore be taken into account. Total and differential cross sections including the full dependence on the top- and bottom-quark masses have been computed up to NNLO accuracy~\cite{Georgi:1977gs,Baur:1989cm,Harlander:2005rq,Chen:2021azt,Czakon:2021yub,Czakon:2023kqm,Niggetiedt:2024nmp,Grazzini:2013mca,Buschmann:2014sia,Melnikov:2016emg,Lindert:2017pky,Caola:2018zye,Bonciani:2022jmb,Czakon:2024ywb}.\\
Nevertheless, for a proof-of-concept validation of \history{}, it is sufficient to consider Higgs-boson production in the heavy-top limit. The required matrix elements are provided in analytic form. Most of them are derived using a streamlined setup based on a \texttt{Mathematica}~\cite{Mathematica14} toolchain built around the package \texttt{Alibrary}~\cite{alibrary}, which employs \texttt{QGRAF}~\cite{Nogueira:1991ex} for diagram generation and \texttt{FORM}~\cite{vanRitbergen:1998pn,Vermaseren:2000nd,Kuipers:2012rf,Ruijl:2017dtg,Ueda:2020wqk} for the evaluation of Dirac and colour algebra. The NNLO contributions involving the interference of the real-virtual corrections with tree-level real-emission amplitudes are taken from Ref.~\cite{Schmidt:1997wr}, while the two-loop form factor entering the double-virtual contribution is adopted from Ref.~\cite{Gehrmann:2010ue}. For the efficient evaluation of the double-real emission amplitudes, we utilise the standalone mode of \texttt{MadGraph5\_aMC@NLO}~\cite{Alwall:2014hca} and link the generated amplitudes to \history{} via an interface to modified, lightweight versions of the \texttt{ALOHA}~\cite{deAquino:2011ub} and \texttt{HELAS}~\cite{Murayama:1992gi} libraries. All matrix elements up to one loop were validated by comparisons with \texttt{OpenLoops2}~\cite{Buccioni:2019sur}.\\
The first benchmark considers the total inclusive cross section in proton–proton collisions at $\sqrt{s}=13.6\,\mathrm{TeV}$. The renormalisation and factorisation scales are chosen as $\mu_\mathrm{F}=\mu_\mathrm{R}=M_H$, with the Higgs boson mass fixed to $M_H=125.2\,\mathrm{GeV}$. The vacuum expectation value entering the effective gluon-gluon-Higgs vertex is derived from Fermi’s constant according to $v=1/\sqrt{\sqrt{2}G_\mathrm{F}}$, with $G_\mathrm{F}=1.1663788\cdot 10^{-5}\,\mathrm{GeV}^{-2}$. For the PDFs, we employ the \texttt{PDF4LHC21\_40}~\cite{PDF4LHCWorkingGroup:2022cjn} set. The results obtained with \history{} are compared in Tab.~\ref{tab:total_xsec_ggh} to those from \sushi{}~\cite{Harlander_2013,Harlander_2017}. We find agreement within the quoted Monte-Carlo uncertainties for both the total cross section and the individual partonic channels. The $gg \to H+X$ channel collects all contributions proportional to the product of two gluon PDFs. The $qg \to H+X$ channel contains contributions proportional to a quark or antiquark PDF multiplied by a gluon PDF, where the gluon may originate from either incoming proton. The $qq^\prime \to H+X$ channel comprises all contributions proportional to the product of two quark and/or antiquark PDFs.\\
\begin{table}[t]
 \renewcommand{\arraystretch}{1.3}
 \centering
 \begin{NiceTabular}{ l | r | r }[color-inside]
 \arrayrulecolor{rwthblue}
    \rowcolor{rwthblue}
	 \phantom{......}\textcolor{white}{channel}\phantom{......} & \phantom{.}\textcolor{white}{\history{} $\mathrm{[fb]}$}\phantom{.} & \phantom{..}\textcolor{white}{\sushi{} $\mathrm{[fb]}$}\phantom{..} \\ \rowcolor{rwthblue50}
	  $pp \to H+X$ & $42518.75(69)$ & $42519.68(12)$ \\ \rowcolor{rwthblue10}
        $gg \to H+X$ & $43585.94(58)$ & $43586.57(12)$ \\ \rowcolor{rwthblue10}
        $qg \to H+X$ & $-1135.68(37)$ & $-1135.403(58)$ \\ \rowcolor{rwthblue10}
        $qq^\prime \to H+X$ & $68.488(18)$ & $68.48029(10)$ \\
 \end{NiceTabular}  
 \caption{Total inclusive NNLO cross section for Higgs production in gluon fusion. Statistical Monte-Carlo uncertainties for the last two digits are quoted in parentheses.}
    \label{tab:total_xsec_ggh} 
\end{table}\noindent
As a second validation test, we compare the Higgs-boson transverse-momentum spectrum obtained with \history{} to the prediction from \nnlojet{}~\cite{huss2025nnlojet} for the setup described above. The distributions are computed using bins of width $\Delta p_{T,H}^\mathrm{bin}=20\,\mathrm{GeV}$ and are shown in Fig.~\ref{fig:Higgs}. Over the entire kinematic range considered, \history{} reproduces the spectrum within the quoted Monte-Carlo uncertainties.\\
\noindent
The successful validation of both the total inclusive cross section and a non-trivial differential observable provides strong evidence for the correctness of our implementation of the NSC algorithm for gluon-induced colour-singlet production.\\
In addition, we performed further consistency checks that are not shown here. These include variations of the factorisation and renormalisation scales, as well as comparisons with \texttt{NNLOCAL}~\cite{DelDuca:2024ovc} for Higgs production and Higgs production in association with a jet at $n_f=0$. The latter process was also compared at $n_f=5$ with \nnlojet{}. The $pp \to Hj+X$ comparisons were carried out at NLO in QCD.\\
Typical runtimes of \history{} on a laptop or MacBook are of the order of five minutes to obtain total cross-section predictions for NNLO $pp \to H+X$ calculations with statistical Monte-Carlo uncertainties at the level of $0.05\%$. The runtime for differential distributions depends strongly on the observable under consideration. The transverse-momentum distribution shown in Fig.~\ref{fig:Higgs} represents a particularly challenging observable, as $p_T$ lies along a diagonal direction of the chosen Monte-Carlo parametrisation and therefore receives only limited benefit from the importance-sampling strategy of the \texttt{VEGAS} algorithm~\cite{Lepage:1977sw,lepage1980vegas}. The distribution was generated using nearly $10^{11}$ events, corresponding to approximately $7200$ core hours on $2.10\,\mathrm{GHz}$ Intel\textsuperscript{\textregistered} Xeon\textsuperscript{\textregistered} Platinum 8468 processors.

\section{Associated Higgs-Boson Production}\label{sec:dy}
\begin{figure}
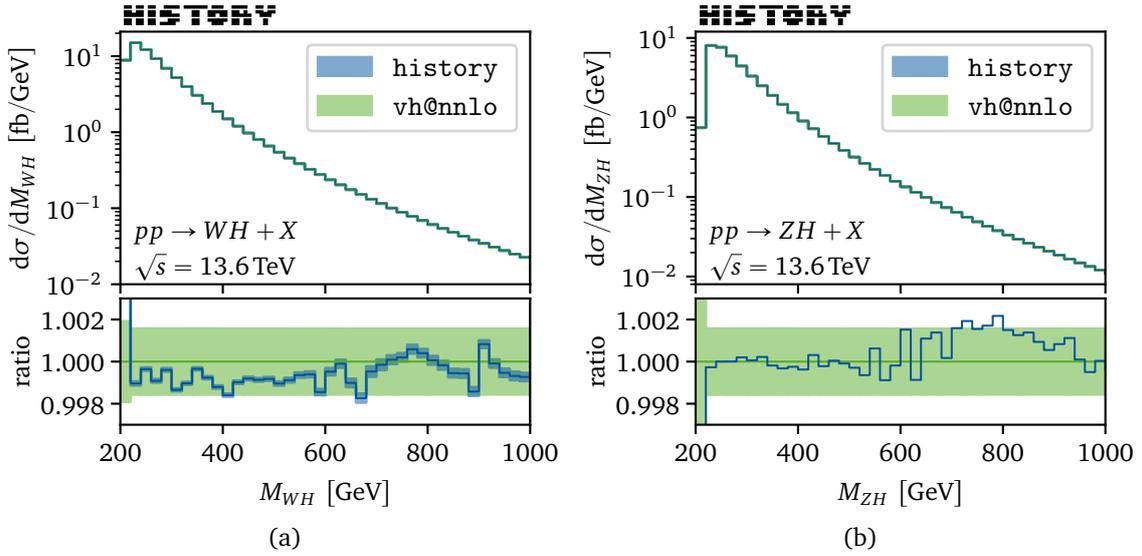

    \begin{subfigure}[b]{0.5\textwidth}
        \input{figures/Fig3a.pgf}
        \caption{}
    \end{subfigure}
    \begin{subfigure}[b]{0.5\textwidth}
        \input{figures/Fig3b.pgf}
        \caption{}
    \end{subfigure}
    \caption{Invariant-mass spectra for (a) $pp \to WH+X$ and (b) $pp \to ZH+X$ at NNLO QCD precision. The upper panels show the spectra computed with \history{} (blue) and \vhnnlo{} (green). The lower panels illustrate their ratios normalised to the \vhnnlo{} distribution. The shaded bands indicate the statistical Monte-Carlo uncertainties.}
	\label{fig:MVH}
\end{figure}

\noindent
The second class of colour-singlet production processes is initiated at Born level by quark-antiquark annihilation. To validate our implementation for this production mechanism, we study the associated production of a Higgs boson with a massive electroweak vector boson $V\in\{W,Z\}$ through the Drell-Yan-like channel, $pp\to V^\ast+X \to VH+X$. At leading order, a light quark $q$ and a light antiquark $\widebar{q}^\prime$ annihilate into a virtual vector boson, which subsequently decays into an on-shell vector boson and a Higgs boson, $q\widebar{q}^\prime\to V^\ast \to VH$. The corresponding Feynman diagram is shown in Fig.~\ref{fig:dy_lo}.\\
The NLO QCD corrections for this process were computed in the early 1990s~\cite{Han:1991ia,Ohnemus:1992bd,Baer:1992vx}, and the NLO electroweak corrections are available for both the total inclusive~\cite{Ciccolini:2003jy} and differential~\cite{Denner:2011id,Denner:2014cla} cross sections. The exact NNLO QCD corrections to the total cross section were obtained more than two decades ago~\cite{Brein_2004}, while fully-differential predictions, including various decay channels of the Higgs and the electroweak gauge bosons, were developed during the past decade~\cite{PhysRevLett.107.152003,Ferrera_2014,FERRERA201551,Boughezal_2016,dur25585,Ferrera_2018,Caola_2018,Gauld_2019,Behring_2020,Bonetti:2025hnb}. More recently, the total cross section has been determined at N\textsuperscript{3}LO accuracy~\cite{Baglio:2022wzu}.\\
To benchmark our implementation, we compare the total inclusive NNLO cross section obtained with \history{} to the prediction of \vhnnlo{}~\cite{Brein:2012ne,Harlander:2018yio}. The required matrix elements contributing at LO and NLO were derived analytically. At NNLO, the interference of the one-loop real-virtual corrections with tree-level single-real emission diagrams was computed analytically using \texttt{Mathematica}~\cite{Mathematica} together with \texttt{Package-X}~\cite{Patel:2015tea,Patel:2016fam}, and the result was numerically cross-checked against \texttt{OpenLoops2}. The two-loop form factor entering the double-virtual corrections was taken from Ref.~\cite{Hamberg:1990np}. To evaluate the double-real emission contributions, the corresponding amplitudes were generated with \texttt{MadGraph5\_aMC@NLO}~\cite{Alwall:2014hca} and linked to \history{} in the same way as for the Higgs-production process.\\
\begin{table}[t]
 \renewcommand{\arraystretch}{1.3}
 \centering
 \begin{NiceTabular}{ l | r | r }[color-inside]
 \arrayrulecolor{rwthblue}
    \rowcolor{rwthblue}
	 \phantom{......}\textcolor{white}{channel}\phantom{......} & \phantom{.}\textcolor{white}{\history{} $\mathrm{[fb]}$}\phantom{.} & \phantom{.}\textcolor{white}{\vhnnlo{} $\mathrm{[fb]}$}\phantom{.} \\ \rowcolor{rwthblue50}
	    $pp \to WH+X$ & $1557.57(16)$ & $1557.56(14)$ \\ \rowcolor{rwthblue10}
        $qq^\prime \to WH+X$ & $1766.52(13)$ & $1766.46(13)$ \\ \rowcolor{rwthblue10}
        $qg \to WH+X$ & $-212.733(95)$ & $-212.679(48)$ \\ \rowcolor{rwthblue10}
        $gg \to WH+q\widebar{q}^\prime$ & $3.77566(86)$ & $3.77584(76)$ \\ \rowcolor{rwthblue50}
        $pp \to ZH+X$ & $848.274(35)$ & $848.269(76)$ \\ \rowcolor{rwthblue10}
        $qq^\prime \to ZH+X$ & $966.131(12)$ & $966.102(71)$ \\ \rowcolor{rwthblue10}
        $qg \to ZH+X$ & $-120.332(10)$ & $-120.346(26)$ \\ \rowcolor{rwthblue10}
        $gg \to ZH+q\widebar{q}$ & $2.51256(15)$ & $2.51319(50)$
 \end{NiceTabular}  
 \caption{Total NNLO cross section for associated Higgs production. Statistical Monte-Carlo uncertainties for the last two digits are quoted in parentheses.}
    \label{tab:total_xsec_dy} 
\end{table}\noindent
Our setup considers a proton–proton collision at $\sqrt{s}=13.6\,\mathrm{TeV}$, with $\mu_\mathrm{F}=\mu_\mathrm{R}=\sqrt{(p_V+p_H)^2}$. We choose the \texttt{PDF4LHC21\_40}~\cite{PDF4LHCWorkingGroup:2022cjn} PDF set. The Higgs mass is fixed at $M_H=125.2\,\mathrm{GeV}$, the vector bosons have the masses $M_W=80.3692\,\mathrm{GeV}$ and $M_Z=91.188\,\mathrm{GeV}$, while the decay widths are set to zero, $\Gamma_W=\Gamma_Z=0$. Electroweak couplings are defined in the $G_\mu$ scheme, where Fermi's coupling constant is taken as $G_\mathrm{F}=1.1663788\cdot 10^{-5}\,\mathrm{GeV}^{-2}$. For $WH$ production, a diagonal CKM matrix is assumed.\footnote{For a consistent comparison of $WH$ production cross sections, bottom-quark PDFs are disabled in \history{} to match the implementation in \vhnnlo{}. In all other respects, the number of light quark flavours is kept at five. In the released version of \history{}, bottom-quark PDFs are enabled again.} The total inclusive cross sections for our validation test are summarised in Tab.~\ref{tab:total_xsec_dy}. Completetly analogous to the comparisons for Higgs production, we decomposed the proton–proton cross section into three distinct partonic channels. The first channel is the cross section induced by two light (anti)quarks, where $qq^\prime$ represents any pair of light quarks and/or antiquarks. The second channel is induced by a light quark or antiquark $q$ and a gluon $g$. The third channel is induced by two gluons. For both $WH$ and $ZH$ production in the Drell-Yan-like production mode, we find the predictions of \history{} and \vhnnlo{} to be fully consistent within Monte-Carlo uncertainties across all channels.\\
To further validate our implementation, we compare the invariant-mass distribution of the final-state $VH$ system, $M_{VH}=\sqrt{(p_V+p_H)^2}$, as obtained with \history{} and \vhnnlo{}. For this comparison, the distribution is computed with \history{} using bins of width $\Delta M_{VH}^\mathrm{bin}=20\,\mathrm{GeV}$. Since \vhnnlo{} is based on an analytic expression for $\mathrm{d}\sigma/\mathrm{d}M_{VH}$ rather than an event-binning approach, its prediction for each bin is obtained by evaluating the differential cross section at a set of $N=20$ equally spaced values of $M_{VH}$ and averaging the results.\ Fig.~\ref{fig:MVH}, using the same setup as in the previous test, shows that \history{} reproduces the invariant-mass spectrum predicted by \vhnnlo{} at the permille level over the entire kinematic range.\\
As an additional consistency check, we repeated the comparison of the total inclusive cross section using different choices of Standard-Model input parameters and renormalisation and factorisation scales, finding no significant deviations or inconsistencies. We also thank the authors of Ref.~\cite{Caola_2019} for providing comparisons of all independently finite integrals for a given parameter setup, which further supported our validation.\\
\noindent
Owing to the increased complexity of the two-particle Born phase space compared to the $pp \to H+X$ case discussed in the previous section, the runtime of \history{} for total NNLO cross-section computations of $pp \to VH+X$ processes increases to typically just under half an hour on a standard laptop for a statistical uncertainty of $0.05\%$.\\
As a final remark, we briefly comment on the numerical stability of the implemented subtraction procedure. All computations presented here are carried out entirely in double precision. This applies to both the matrix elements and the local subtraction terms that regulate IR divergences in real-emission contributions. To ensure finite and stable results, a numerical cutoff is applied only to angular variables in integrands involving double-real emission matrix elements. In the double-collinear partitions, we require $\eta_{ij}>10^{-12}$, while in the triple-collinear partitions we impose $\eta_{ij}>10^{-14}$. Exceptions are sectors $(b)$ and $(d)$, where cuts of $\eta_{45}>10^{-7}$ and $\eta_{45} > 10^{-7.5}$ are applied for integrands with and without energy ordering, respectively.
\section{Conclusions}\label{sec:conclustions}

In this work, we have presented \history{}, a new parton-level Monte-Carlo program that implements the fully-local nested soft-collinear subtraction scheme for the computation of fully-differential cross sections for colour-singlet production processes in hadronic collisions up to NNLO accuracy in QCD. The implementation of the subtraction algorithm is process independent, such that only the corresponding matrix elements need to be supplied for a given process. This structure makes \history{} a flexible and extensible framework for high-precision QCD computations of colour-singlet observables.\\
The correctness and robustness of the implementation have been validated through comprehensive benchmark studies. Since colour-singlet production at Born level proceeds either via gluon fusion or quark-antiquark annihilation, we considered one representative process from each class. For gluon-induced production, Higgs boson production in the heavy-top limit was studied and compared against independent calculations, including \sushi{} and \nnlojet{}. For quark-induced production, associated Higgs production via the Drell-Yan-like Higgs-Strahlung mechanism was validated against \vhnnlo{}. In all cases, excellent agreement was observed for both inclusive and differential observables, confirming the reliability of the framework.\\
With its process-independent subtraction algorithm, modular matrix-element interface, and demonstrated numerical stability, \history{} provides a solid foundation for future NNLO studies of colour-singlet production processes. The public release of the code together with this publication aims to facilitate further developments and applications, including the extension to additional processes and the incorporation of more sophisticated amplitude providers.\\
We anticipate that \history{} will serve as a useful tool for precision phenomenology at the LHC and future hadron colliders, with the long-term goal of establishing a general semi-automated framework for NNLO computations based on local subtraction techniques.

\section*{Acknowledgements}
We thank Micha\l{} Czakon, Terry Generet, Robert Harlander, and Hua-Sheng Shao for valuable discussions. We are particularly grateful to Raoul Röntsch for his kind advice and for providing independent cross-checks to validate our results. We also thank Robert Harlander for carefully proofreading the manuscript and for his helpful comments.\\[-2\baselineskip]
\paragraph{Funding information:} This research was supported by the Deutsche Forschungsgemeinschaft (DFG, German Research Foundation) under grant 400140256 -- \textit{GRK 2497: The physics of the heaviest particles at the LHC} and grant 396021762 -- \textit{TRR 257: P3H – Particle Physics Phenomenology after the Higgs Discovery}. S.Y.K.\@ receives funding from the Federal Ministry of Education and Research (BMBF) under grant number 05H21PACCA. The work of L.S.\@ was supported by ERC grant 101041109 -- \textit{BOSON}. Views and opinions expressed are however those of the authors only and do not necessarily reflect those of the European Union or the European Research Council Executive Agency. Neither the European Union nor the granting authority can be held responsible for them. Computations were performed with computing resources granted by RWTH Aachen University under project rwth1826.

\begin{appendix}
\numberwithin{equation}{section}
\section{Installation and Execution}\label{sec:history}\noindent
The source code of \history{} is publicly available and can be obtained from the GitLab repository  
\href{https://gitlab.com/history-nnlo/history}{https://gitlab.com/history-nnlo/history}. Before compiling the executable, all required external dependencies listed below must be installed and properly linked to ensure a successful build.

\subsection{Dependencies}\noindent
The following external dependencies are required to install \history{}:

\begin{itemize}

\item {\fontfamily{lmtt}\fontseries{b}\selectfont LHAPDF}  
Access to standardised PDF sets is provided through an interface to the \texttt{LHAPDF} library~\cite{Andersen:2014efa,Buckley_2015}. It can be found at \href{https://lhapdf.hepforge.org}{https://lhapdf.hepforge.org}. After installation, the environment variable \texttt{LHAPDF\_DATA\_PATH} must point to the directory containing the PDF sets to ensure that \history{} can locate the corresponding grids at runtime.

\item {\fontfamily{lmtt}\fontseries{b}\selectfont Cuba} Numerical integrations are performed using the \texttt{VEGAS} algorithm~\cite{Lepage:1977sw,lepage1980vegas}, as efficiently implemented in the \texttt{Cuba} library~\cite{Hahn_2005,hahn2014concurrent}. The library can be downloaded from \href{https://feynarts.de/cuba}{https://feynarts.de/cuba}.

\item {\fontfamily{lmtt}\fontseries{b}\selectfont Polylogarithm} The evaluation of classical polylogarithms is handled via the library \texttt{Polylogarithm}~\cite{polylogarithm}. The library provides fast and numerically stable implementations suitable for high-precision applications and is publicly available from the GitHub repository \href{https://github.com/Expander/polylogarithm}{https://github.com/Expander/polylogarithm}.

\item {\fontfamily{lmtt}\fontseries{b}\selectfont LoopTools} Scalar one-loop integrals appearing in some amplitudes are evaluated using \texttt{LoopTools}~\cite{Hahn_1999}, which provides numerical implementations of the Passarino-Veltman integrals. The library can be obtained from \href{https://feynarts.de/looptools}{https://feynarts.de/looptools}.

\item {{\fontfamily{lmtt}\fontseries{b}\selectfont FastJet} \textbf{(optional)}} 
The computation of colour-singlet production in association with two jets at LO or with one jet at NLO is supported in \history{} through the use of jet clustering algorithms. The code includes built-in implementations of the $k_t$~\cite{Catani:1993hr,Ellis:1993tq}, anti-$k_t$~\cite{Cacciari:2008gp}, and Cambridge/Aachen~\cite{Dokshitzer:1997in,Wobisch:1998wt} jet algorithms. Optionally, \texttt{FastJet}~\cite{Cacciari:2011ma,Cacciari:2005hq} can be linked in order to employ its implementations of these algorithms. \texttt{FastJet} is available from \href{https://fastjet.fr}{https://fastjet.fr}. To compile \history{} with \texttt{FastJet} support, the flag \texttt{USE\_FJ=1} must be set in the \texttt{config} file.

\end{itemize}

\subsection{Installation}\noindent
After installing the dependencies, the compilation of \history{} can be performed. All necessary paths must be properly configured and, if external libraries are not installed in default locations, the corresponding paths should be specified in the \texttt{config} file located in the main directory of \history{}. Compilation is carried out via running the command \texttt{make history}. To accelerate the build process, parallel compilation can be enabled using the \texttt{-j} option. For example, compilation with eight parallel jobs can be invoked through
\begin{bashlisting}[\hspace{9pt}\footnotesize{shell}]
 make history -j8
\end{bashlisting}
The compilation time is typically less than one minute, even when performed on a single core. Upon successful completion, the executable \texttt{history} is generated in the directory \texttt{history/bin}.\\
Recompilation may become necessary, for example, after adding a new user-defined observable in \texttt{history/source/user.F90} or when linking against a different version of an external dependency. In such cases, the build directory should be cleaned before regenerating the executable:
\begin{bashlisting}[\hspace{9pt}\footnotesize{shell}]
 make clean && make history -j8
\end{bashlisting}

\subsection{Execution}\noindent
Execution of \history{} is performed from the command line and accepts up to two arguments. The first argument specifies the input file (see Appendix~\ref{sec:input}), while the second optional argument defines the output file. If no output file is provided, a file with the same name as the input file and the extension \texttt{.out} is generated automatically. In this case, the output is written to the same directory as the input file. For example, executing \history{} from the main directory with the input file \texttt{inputfile.in} located in \texttt{history/input} and writing the results to \texttt{outputfile.out} in \texttt{history/output} can be achieved via
\begin{bashlisting}[\hspace{9pt}\footnotesize{shell}]
 ./bin/history ./input/inputfile.in ./output/outputfile.out
\end{bashlisting}
Histograms are stored separately for each subprocess in the same directory as the output file. The corresponding file names follow the convention \texttt{outputfile\_proc\_obs.hist}, where \texttt{proc} identifies the computed subprocess and \texttt{obs} denotes the observable. If a file with the specified output name already exists, it will be overwritten.

\section{Input Files}\label{sec:input}\noindent
All settings for running \history{} are defined in a single input file and inputs are controlled through SLHA-like input blocks~\cite{Skands_2004,Allanach_2009}. Each parameter belongs to a specific input block and is associated with an index of this block. A block can be initialised by the keyword \texttt{Block} followed by the name of the block. In the subsequent lines the parameters can be set by specifying the index of the parameter at the beginning of the line and then giving a value to the parameter, whereby the value has to be separated from the index by a least one space. A simple example for such an input block is the block \texttt{COLLIDER} which controls the input parameters associated to the particle collider. The collider type is associated with the index \texttt{1} within this block and the collisions hadronic centre-of-mass energy is associated with the index \texttt{2}. In order to set the collider type to a proton–proton collider (which corresponds to the value \texttt{1} and is currently the only option) and the collision energy to $\sqrt{s}=13.6\,\mathrm{TeV}$, the block has to have the format:
\begin{lstlisting}[language=fortran, style=slha]
Block COLLIDER
   1                 1   # collider type [1:proton-proton]
   2    1.36000000E+04   # hadronic centre-of-mass energy [GeV]
\end{lstlisting}
Comments start with the symbol \# and everything following this symbol will be ignored. A complete input file consists of multiple blocks, an example is given in Listing~\ref{lst:inputfile} and some predefined input files can be found in the directory \texttt{history/input/}.\\
With a growing process library and an increasing number of supported features, the structure of the input file might change continuously. Hence, we do not want to go too much into the details of the current version's input parameters. Instead, in order to have an always up-to-date documentation, we point to the regularly maintained user manual that comes along with the distribution of the source code which can be found in the directory \texttt{history/doc/}.
\begin{lstlisting}[language=fortran, style=slha, caption={Example input file for \history{}.}, label={lst:inputfile},float,floatplacement=H]
Block VERSION
   1             1.0.0   # input file version
Block SINGLEHIGGS
   1                 2   # order [-1:off, 0:LO,1:NLO,2:NNLO]
   2                 1   # heavy-top limit (HTL) [0:no, 1:yes]
   3                 1   # Born-improved HTL [0:no, 1:yes]
Block HIGGSSTRAHLUNG
   1                23   # final state [23:ZH, 24:WH]
   2                 2   # order [-1:off, 0:LO,1:NLO,2:NNLO]
Block COLLIDER
   1                 1   # collider type [1:proton-proton]
   2    1.36000000E+04   # hadronic centre-of-mass energy [GeV]
Block PDF
   1      PDF4LHC21_40   # PDF set name
   2                 0   # PDF set member
Block SCALE
   1                 1   # muR [0:fixed, 1:dynamic]
   2                 1   # muF [0:fixed, 1:dynamic]
   3    1.00000000E+00   # multiplicative factor for muR
   4    1.00000000E+00   # multiplicative factor for muF
   5    9.11880000E+01   # fixed muR value (if enabled) [GeV]
   6    9.11880000E+01   # fixed muF value (if enabled) [GeV]
Block SMPARAMETER
   1    1.16637880E-05   # Fermi constant [GeV^-2]
Block MASS
   5    4.78000000E+00   # bottom-quark pole mass [GeV]
   6    1.72570000E+02   # top-quark pole mass [GeV]
  23    9.11880000E+01   # Z-boson mass [GeV]
  24    8.03692000E+01   # W-boson mass [GeV]
  25    1.25200000E+02   # Higgs mass [GeV]
Block DECAY
  23    2.49550000E+00   # decay width of the Z boson [GeV]
  24    2.08500000E+00   # decay width of the W boson [GeV]
Block DISTRIBUTION
   1             PT(1)   # observable [PT: transverse momentum]  
   2    0.00000000E+00   # lower edge of histogram range [GeV]
   3    1.00000000E+03   # upper edge of histogram range [GeV]
   4    2.00000000E+01   # histogram bin width [GeV]
   5           1000000   # number of events for histogramming
Block MONTECARLO
   1                 1   # integration algorithm [1:VEGAS]
   2             28236   # random seed
   3                 8   # number of parallel CPU threads
   4    5.00000000E-05   # target relative precision
   5            100000   # minimum number of events
   6           1000000   # maximum number of events
   7             10000   # events in first iteration
   8                 0   # increment of events per iteration
\end{lstlisting}

\section{Splitting Functions}\label{sec:splitting}\noindent
The space-like Altrarelli-Parisi splitting functions that are required for NNLO QCD computations are
\begin{equation}
	\begin{aligned}
		\hat{P}_{q_i q_j}(z) &= \delta_{ij} \hat{P}^{(0)}_{qq}(z) + \dfrac{\alpha_s(\mu)}{2\pi} \hat{P}^{(1)}_{q_i q_j}(z) + \mathcal{O}\left(\alpha_s^2\right)\,, \\
		\hat{P}_{qg}(z) &= \hat{P}^{(0)}_{qg}(z) + \dfrac{\alpha_s(\mu)}{2\pi} \hat{P}^{(1)}_{qg}(z) + \mathcal{O}\left(\alpha_s^2\right)\,,\\
		\hat{P}_{gq}(z) &= \hat{P}^{(0)}_{gq}(z) + \dfrac{\alpha_s(\mu)}{2\pi} \hat{P}^{(1)}_{gq}(z) + \mathcal{O}\left(\alpha_s^2\right)\,, \\
		\hat{P}_{gg}(z) &= \hat{P}^{(0)}_{gg}(z) + \dfrac{\alpha_s(\mu)}{2\pi} \hat{P}^{(1)}_{gg}(z) + \mathcal{O}\left(\alpha_s^2\right)\,.
	\end{aligned}
\end{equation}
The Altarelli–Parisi splitting functions can be further decomposed into
\begin{equation}
	\hat{P}^{(k)}_{ij}(z) = \hat{P}^{(k)}_{ij,\mathrm{R}}(z) +\hat{P}^{(k)}_{ij,\delta}\, \delta(1-z)\,.
\end{equation}
The LO splitting functions were computed in Ref.~\cite{ALTARELLI1977298}. The regular terms read
\begin{equation}
	\begin{aligned}
		\hat{P}^{(0)}_{qq,\mathrm{R}}(z) &= C_F \left( \dfrac{2}{\lbrack 1-z \rbrack_{+}}-(1+z) \right)\,,\\
		\hat{P}^{(0)}_{qg,\mathrm{R}}(z) &= T_R \left( z^2+(1-z)^2 \right)\,,\\
		\hat{P}^{(0)}_{gq,\mathrm{R}}(z) &= C_F \left( \dfrac{1+(1-z)^2}{z} \right)\,,\\
		\hat{P}^{(0)}_{gg,\mathrm{R}}(z) &= 2\, C_A \left( \dfrac{1}{\lbrack 1-z \rbrack_+}+\dfrac{1}{z}+z\,(1-z)-2 \right)\,,
	\end{aligned}
\end{equation}
and the terms proportional to the $\delta$-distribution are
\begin{equation}
	\begin{aligned}
		\hat{P}^{(0)}_{qq,\delta} &= \gamma_q = \dfrac{3}{2} C_F\,,\\
		\hat{P}^{(0)}_{qg,\delta} &= 0\,,\\
		\hat{P}^{(0)}_{gq,\delta} &= 0\,,\\
		\hat{P}^{(0)}_{gg,\delta} &= \gamma_g = \beta_{0} = \dfrac{11}{6}C_A - \dfrac{2}{3}T_Rn_f\,.
	\end{aligned}
\end{equation}
The NLO splitting functions are taken from Refs.~\cite{ellis_stirling_webber_1996,Czakon_2015}. It is convenient to distinguish the flavour singlet ($\mathrm{S}$) and non-singlet ($\mathrm{V}$) contributions for a quark decaying into a quark,
\begin{equation}\label{eq:pqq_NLO}
	\begin{aligned}
		\hat{P}^{(1)}_{q_i q_j}(z) &= \delta_{ij} \left(\hat{P}^{\mathrm{V}(1)}_{qq,\mathrm{R}}(z) + \hat{P}^{(1)}_{qq,\delta}\, \delta(1-z) \right) + \hat{P}^{\mathrm{S}(1)}_{qq,\mathrm{R}}(z)\,, \\
		\hat{P}^{(1)}_{q_i \widebar{q}_j}(z) &= \delta_{ij} \hat{P}^{\mathrm{V}(1)}_{q\widebar{q},\mathrm{R}}(z) + \hat{P}^{\mathrm{S}(1)}_{qq,\mathrm{R}}(z)\,.
	\end{aligned}
\end{equation} 
For a quark decaying into a quark, the non-singlet term is given by
\begin{equation}
	\begin{aligned}
		\hat{P}^{\mathrm{V}(1)}_{qq,\mathrm{R}}(z) &= C_F^2 \left\lbrack-\left(2\ln(z)\ln(1-z)+\dfrac{3}{2}\ln(z)\right)p_{qq}(z)-\left(\dfrac{3}{2}+\dfrac{7}{2}z\right)\ln(z)\right. \\
		&\quad \left. -\dfrac{1}{2}(1+z)\ln^2(z)-5(1-z) \right\rbrack +C_A C_F \left\lbrack \left(\dfrac{1}{2}\ln^2(z)+\dfrac{11}{6}\ln(z) \right.\right.\\
		&\quad \left.\left. +\dfrac{67}{18}-\dfrac{\pi^2}{6}\right) p_{qq}(z)+(1+z)\ln(z) + \dfrac{20}{3}(1-z) \right\rbrack + C_F T_R n_f \\
		&\quad \times \left\lbrack -\left(\dfrac{2}{3}\ln(z)+\dfrac{10}{9}\right)p_{qq}(z) - \dfrac{4}{3}(1-z)\right\rbrack\,.
	\end{aligned}
\end{equation}
For a quark decaying into an antiquark the non-singlet term is
\begin{equation}
	\hat{P}^{\mathrm{V}(1)}_{q\widebar{q},\mathrm{R}}(z) = C_F \left(C_F-\dfrac{1}{2}C_A\right)\bigl\lbrack 2\,p_{qq}(-z)\, S_2(z)+2\,(1+z)\ln(z)+4\,(1-z) \bigr\rbrack\,.
\end{equation}
The singlet contribution is the same in both expressions of Eq.~(\ref{eq:pqq_NLO}),
\begin{equation}
	\hat{P}^{\,\mathrm{S}\,(1)}_{qq,\mathrm{R}}(z) = C_F T_R \left\lbrack -2+\dfrac{20}{9z}+6z-\dfrac{56}{9}z^2+\left(1+5z+\dfrac{8}{3}z^2\right)\ln(z)-(1+z)\ln^2(z) \right\rbrack\,.
\end{equation}
The NLO splitting function for a gluon decaying into a quark is
\begin{equation}
	\begin{aligned}
		\hat{P}^{(1)}_{qg,\mathrm{R}}(z) &= \dfrac{1}{2}C_FT_R \biggl\lbrack 4-9z-(1-4z)\ln(z)-(1-2z)\ln^2(z)+4\ln(1-z) \\
		&\quad +\left(2\ln^2\left(\dfrac{1-z}{z}\right)-4\ln\left(\dfrac{1-z}{z}\right)-\dfrac{2\pi^2}{3}+10\right)p_{qg}(z)\biggr\rbrack \\
		&\quad +\dfrac{1}{2}C_A T_R \biggl\lbrack \dfrac{182}{9}+\dfrac{14}{9}z+\dfrac{40}{9z}+\left(\dfrac{136}{3}z-\dfrac{38}{3}\right)\ln(z)-4\ln(1-z) \\
		&\quad -(2+8z)\ln^2(z) + 2\,p_{qg}(-z)\,S_2(z)+\left(-\ln^2(z)+\dfrac{44}{3}\ln(z) \right. \\
		&\quad\left. -2\ln^2(1-z)+4\ln(1-z)+\dfrac{\pi^2}{3}-\dfrac{218}{9} \right) p_{qg}(z) \biggr\rbrack\,.
	\end{aligned}
\end{equation}
The NLO splitting function for a quark splitting into a gluon is
\begin{equation}
	\begin{aligned}
		\hat{P}^{(1)}_{gq,\mathrm{R}}(z) &= C_F^2 \Bigl\lbrack -\dfrac{5}{2}-\dfrac{7}{2}z+\left(2+\dfrac{7}{2}z\right)\ln(z) - \left(1-\dfrac{1}{2}z\right)\ln^2(z)-2z\ln(1-z) \\
		&\quad -\left(3\ln(1-z)+\ln^2(1-z)\right) p_{gq}(z) \Bigr\rbrack +C_A C_F \left\lbrack \dfrac{28}{9}+\dfrac{65}{18}z+\dfrac{44}{9}z^2 \right. \\
		&\quad \left. -\left(12+5z+\dfrac{8}{3}z^2\right)\ln(z)+(4+z)\ln^2(z)+2z\ln(1-z) \right. \\
		&\quad\left. +S_2(z)\,p_{gq}(-z) +\left(\dfrac{1}{2}-2\ln(z)\ln(1-z)+\dfrac{1}{2}\ln^2(z)+\dfrac{11}{3}\ln(1-z) \right.\right. \\
		&\quad\left.\left. +\ln^2(1-z)-\dfrac{\pi^2}{6} \right)p_{gq}(z)\right\rbrack + C_F T_R n_f \biggl\lbrack -\dfrac{4}{3}z-\left(\dfrac{20}{9}+\dfrac{4}{3}\ln(1-z)\right) \\
		&\quad \times p_{gq}(z) \biggr\rbrack\,.
	\end{aligned}
\end{equation}
The NLO splitting function for a gluon splitting into a gluon is
\begin{equation}
	\begin{aligned}
		\hat{P}^{(1)}_{gg,\mathrm{R}}(z) &= C_A^2 \left\lbrack \dfrac{27}{2}(1-z)+\dfrac{67}{9}\left(z^2-\dfrac{1}{z}\right) -\left(\dfrac{25}{3}-\dfrac{11}{3}z+\dfrac{44}{3}z^2\right)\ln(z) \right. \\
		&\quad \left. +4(1+z)\ln^2(z)+2\,p_{gg}(-z)\,S_2(z) +\left(\dfrac{67}{9}-4\ln(z)\ln(1-z) \right.\right. \\
		&\quad \left.\left. +\ln^2(z)-\dfrac{\pi^2}{3}\right)p_{gg}(z) \right\rbrack +C_F T_R n_f \biggl\lbrack -16+8z+\dfrac{20}{3}z^2 + \dfrac{4}{3z} \\
		&\quad -(6+10z)\ln(z)-2(1+z)\ln^2(z) \biggr\rbrack + C_A T_R n_f \biggl\lbrack 2(1-z) \\
		&\quad +\dfrac{26}{9}\left(z^2-\dfrac{1}{z}\right)-\dfrac{4}{3}(1+z)\ln(z)-\dfrac{20}{9}p_{gg}(z) \biggr\rbrack\,.
	\end{aligned}
\end{equation}
In the previous expressions we have introduced the symbols
\begin{equation}
	S_2(z) = -2\,\mathrm{Li}_2(-z)+\dfrac{1}{2}\ln^2(z)-2\ln(z)\ln(1+z)-\dfrac{\pi^2}{6}
\end{equation}
and
\begin{equation}
	\begin{aligned}
		p_{qq}(z) &= \dfrac{2}{\lbrack 1-z \rbrack_+}-1-z \,,\\
		p_{qg}(z) &= z^2+(1-z)^2\,,\\
		p_{gq}(z) &= \dfrac{1+(1-z^2)}{z}\,,\\
		p_{gg}(z) &= \dfrac{1}{\lbrack 1-z \rbrack_+} + \dfrac{1}{z}-2+z(1-z)\,.
	\end{aligned}
\end{equation}
If the argument of $p_{qq}$ or $p_{gg}$ has a negative sign, the plus-distributions are removed,
\begin{equation}
	\begin{aligned}
		p_{qq}(-z)& = \dfrac{2}{1+z}-1+z\,, \\
		p_{gg}(-z) &= \dfrac{1}{1+z}-\dfrac{1}{z}-2-z(1+z)\,.
	\end{aligned}
\end{equation}
The $\delta$-terms in the soft limit read
\begin{equation}
	\begin{aligned}
		\hat{P}^{(1)}_{qq,\delta} &= C_F^2\left(\dfrac{3}{8}-\dfrac{\pi^2}{2}+6\zeta_3\right) + C_FC_A\left(\dfrac{17}{24}+\dfrac{11\pi^2}{18}-3\zeta_3\right)-C_FT_Rn_f\left(\dfrac{1}{6}+\dfrac{2\pi^2}{9}\right)\,,\\
		\hat{P}^{(1)}_{qg,\delta} &= 0\,,\\
		\hat{P}^{(1)}_{gq,\delta} &= 0\,,\\
		\hat{P}^{(1)}_{gg,\delta} &= C_A^2\left(\dfrac{8}{3}+3\zeta_3\right)-C_FT_Rn_f-\dfrac{4}{3}C_AT_Rn_f\,.
	\end{aligned}
\end{equation}

\end{appendix}





\newpage
\bibliography{SciPost_BiBTeX_File.bib}


\end{document}